\documentclass[trackchanges]{aastex7}

\usepackage{float}
\usepackage{graphicx}
\usepackage{cases}
\usepackage{natbib}
\usepackage{multirow} 
\usepackage{epstopdf}
\shorttitle{Three-dimensional evolution of a solar filament}
\shortauthors{Zhang et al.}

\begin{document}
	
	\title{Three-dimensional evolution of a solar filament with multipoint observations}
	
	\correspondingauthor{Qingmin Zhang}
	\email{zhangqm@pmo.ac.cn}
	
	\author[0000-0003-4078-2265]{Qingmin Zhang}
	\affiliation{Purple Mountain Observatory, Chinese Academy of Sciences, Nanjing 210023, People's Republic of China}
	\email{zhangqm@pmo.ac.cn}
	
	\author[0000-0003-4787-5026]{Jun Dai}
	\affiliation{Purple Mountain Observatory, Chinese Academy of Sciences, Nanjing 210023, People's Republic of China}
	\affiliation{Astronomical Observatory, Kyoto University, Sakyo, Kyoto, Japan}
	\email{daijun@pmo.ac.cn}
	
	\author[0000-0001-8402-9748]{Beili Ying}
	\affiliation{Purple Mountain Observatory, Chinese Academy of Sciences, Nanjing 210023, People's Republic of China}
	\email{yingbl@pmo.ac.cn}
	
	\author[0000-0002-1190-0173]{Ye Qiu}
	\affiliation{Institute of Science and Technology for Deep Space Exploration, Suzhou Campus, Nanjing University, Suzhou 215163, People's Republic of China}
	\email{qiuye@smail.nju.edu.cn}
	
	\author[0000-0003-4655-6939]{Li Feng}
	\affiliation{Purple Mountain Observatory, Chinese Academy of Sciences, Nanjing 210023, People's Republic of China}
	\email{lfeng@pmo.ac.cn}
	
	\author[0000-0001-7693-4908]{Chuan Li}
	\affiliation{School of Astronomy and Space Science, Nanjing University, Nanjing 210023, People's Republic of China}
	\email{lic@nju.edu.cn}
	
	\author[0000-0001-5705-661X]{Hongqiang Song}
	\affiliation{School of Space Science and Technology, Shandong University, Weihai, Shandong 264209, People's Republic of China}
	\email{hqsong@sdu.edu.cn}
	
	\author[0000-0002-3341-0845]{Yue Zhou}
	\affiliation{Purple Mountain Observatory, Chinese Academy of Sciences, Nanjing 210023, People's Republic of China}
	\email{yuezhou@pmo.ac.cn}
	
	\author[]{Zongyi Li}
	\affiliation{Purple Mountain Observatory, Chinese Academy of Sciences, Nanjing 210023, People's Republic of China}
	\email{zyli@pmo.ac.cn}

\begin{abstract}
In this paper, we first devise a geometrical model, featuring a torus-like flux rope based on the shape of 3DCORE model.
The global shape of the torus is an ellipse, while the cross sections are circular along the torus.
The thinnest point is located between the Sun center and photosphere. Deflections and inclination are considered as well.
Using multiwavelength observations from perspectives of Earth, Ahead-STEREO (STA), and Solar Orbiter,
we apply the model to three-dimensional (3D) reconstructions and tracking of the filament eruption, 
which was associated with a flare and a coronal mass ejection (CME) on 2024 October 8.
The morphology, direction, and true velocity ($\sim$433 km\,s$^{-1}$) of the eruptive filament are obtained.
It is found that the filament propagates nonradially, deflecting slightly eastward by $\sim$10$\degr$ and significantly southward by $\sim$40$\degr$.
Trajectory of the filament in the ecliptic plane reveals that the filament moves toward STA.
The true direction of the eruptive filament using imaging and spectral observations is mutually verified by 3D reconstructions.
The heliocentric distance of the filament increases from $\sim$1.68 to $\sim$2.94\,$R_{\sun}$ within 35 minutes.
Based on the results of 3D reconstructions, the true speed of the CME leading edge is evaluated to be 1046$-$1145 km\,s$^{-1}$.
\end{abstract}
	
\keywords{Sun prominences --- Sun flares ---  Solar coronal mass ejections}

\section{Introduction} \label{intro}
Solar filaments are cool and dense plasmas suspending in the hot corona \citep[see reviews][and references therein]{par14,lia25}.
The gravity of a filament is balanced by the magnetic tension force of magnetic dips in sheared arcades \citep{liu12,zqm12} or flux ropes \citep{aul98,xia14,chd25}.
Filaments could keep stable for a few weeks or even months before fading out or escaping from the corona as a result of eruptions \citep{dai21}.
Successful filament eruptions are capable of producing jets \citep{ste16}, flares \citep{shi95,jan15}, and coronal mass ejections \citep[CMEs;][]{for06,chc25}.
Filaments usually serve as the bright cores of CMEs with three-part structures \citep{shq22}.

Observations of filaments and CMEs have a long history \citep{mu25}. However, most of them are from a single perspective, obtaining images at the plane of the sky (POS).
The launch of the Solar TErrestrial RElations Observatory \citep[STEREO;][]{kai08} in October 2006 revolutionized the way we observe the Sun. 
It is composed of two identical satellites with increasing separations, one ahead of the Earth (STA) and the other behind (STB).
Cooperating with other spacecrafts positioning along the Sun-Earth line, such as the Atmospheric Imaging Assembly \citep[AIA;][]{lem12} 
on board the Solar Dynamics Observatory \citep[SDO;][]{pes12} and the Large Angle Spectroscopic Coronagraph \citep[LASCO;][]{bru95} 
on board the Solar and Heliospheric Observatory (SOHO), observations from two or three different perspectives allow for three-dimensional (3D) reconstructions 
of filaments and CMEs, which are crucial to understanding their geometrical and kinematical evolutions.
The launch of the Solar Orbiter \citep[SolO;][]{mu20} in February 2020 provides a new and closer perspective of the Sun \citep{ant21,gyh25,lxh25,dyd26}.

A large number of techniques have been developed to perform 3D reconstructions \citep{mie10,kay24}. 
The tie-pointing or triangulation method utilizes simultaneous observations from two perspectives \citep{in06,bem09,lie09,tho12}.
Identification of the same feature in these images is required. 
The reconstruction is performed using the SolarSoft (SSW\footnote{https://www.lmsal.com/solarsoft/ssw\_install\_howto.html}) routine \texttt{scc\_measure.pro} \citep{chc25}.
\citet{feng12} devised a 3D mask fitting reconstruction method and applied it to the CME on 2010 August 7.
It is noted that these methods do not make assumptions about the morphology of filaments or CMEs.
The forward-modeling approach, on the other hand, assume that CMEs have specific shapes, 
such as icecream cones \citep{zhao02,mich03,mich06} or flux ropes \citep{lep90,zur06}.
In the graduated cylindrical shell (GCS) model \citep{the06}, the flux rope is composed of two identical legs and a central circulus, resembling a croissant.
The model has been widely used in the 3D reconstructions of CMEs \citep{kw14,mo14,mo15,ver23} 
and recently applied to the reconstruction of an eruptive prominence \citep{zqm25}.
\citet{is16} developed a novel Flux Rope in 3D (FRi3D) model to investigate the shape of a CME that undergoes various kinds of deformations, 
such as expansion, rotation, and skewing. Magnetic field is considered in their model.
\citet{mo18} created a new semiempirical 3D COronal Rope Ejection (3DCORE) model for simulating the propagation of CME flux ropes.
The shape of the flux rope is a self-similarly expanding torus-like structure, which is always attached to the Sun (see their Fig. 1).
The structure includes an embedded Gold-Hoyle-like magnetic field with a time-invariant twist number.
The whole model contains the interplanetary propagation, the measurement by a synthetic spacecraft, 
and production of a geomagnetic index from the simulation \citep{wei21a,wei21b,rud24}.

Considering that plenty of eruptions are nonradial in their early evolutions \citep{chen22,hhd26}, \citet{zqm21} put forward a revised cone model.
The tip of the cone is at the source region of the eruption rather than the Sun center. The direction of propagation is characterized by two deflection angles
($\theta_1$ and $\phi_1$) relative to the local vertical. 
The model has been applied to the reconstructions of CMEs \citep{zqm22} as well as prominences \citep{zqm24,lsy25b}.
Recently, \citet{zyj26} investigated two successive CMEs originating from NOAA active region (AR) 12994 and propagating nonradially on 2022 April 15. 
Reconstructions using multipoint observations and the revised cone model indicate that the CMEs head for Mercury at speeds of $\sim$636 and $\sim$696 km\,s$^{-1}$.
Similarly, \citet{zqm23} modified the GCS model by placing the footpoints of two legs at the source region and applied it to the reconstruction of an eruptive 
prominence on 2022 September 23. Deflection, expansion, and acceleration of CMEs and prominences are satisfactorily reproduced with these revised models.

Although there are substantial observations and numerical simulations \citep{lin00,pats10,cx20,xing25,zqm26}, 
the early kinetic and morphological evolutions of solar eruptions are still unclear.
As mentioned above, plenty of models were proposed to explore the 3D morphology and kinetic evolutions of filaments as well as CMEs.
Specifically, the 3DCORE model has been successfully applied to 3D reconstructions of CMEs and interplanetary CMEs (ICMEs).
However, it has never been used to the reconstructions of filaments, especially in their early evolutions.
On 2024 October 8, a long filament erupted from AR 13847 (S27W38) in the southern hemisphere, generating three flare ribbons and a CME.
The nonradial filament eruption was observed by a fleet of spacecrafts from three perspectives, providing a splendid chance to perform 3D reconstructions of the filament.
In this paper, we introduce a geometrical model similar to the shape of 3DCORE and add two deflection angles as in the revised cone model.
The model is immediately applied to 3D reconstructions and tracking of the filament on 2024 October 8.
It is found that the filament propagates nonradially with slight eastward deflection and considerable southward deflection in the early evolution. 
Meanwhile, the filament and associated CME head for STEREO-A instead of Earth in the ecliptic plane, which is instructive for space weather forecast.
The paper is organized as follows. The geometrical model is described in Section~\ref{model}.
Data analysis of multipoint observations of the filament eruption are presented in Section~\ref{obs}.
Finally, a brief summary and discussions are given in Section~\ref{sum}.

\begin{figure} 
   \includegraphics[width=0.45\textwidth,clip=]{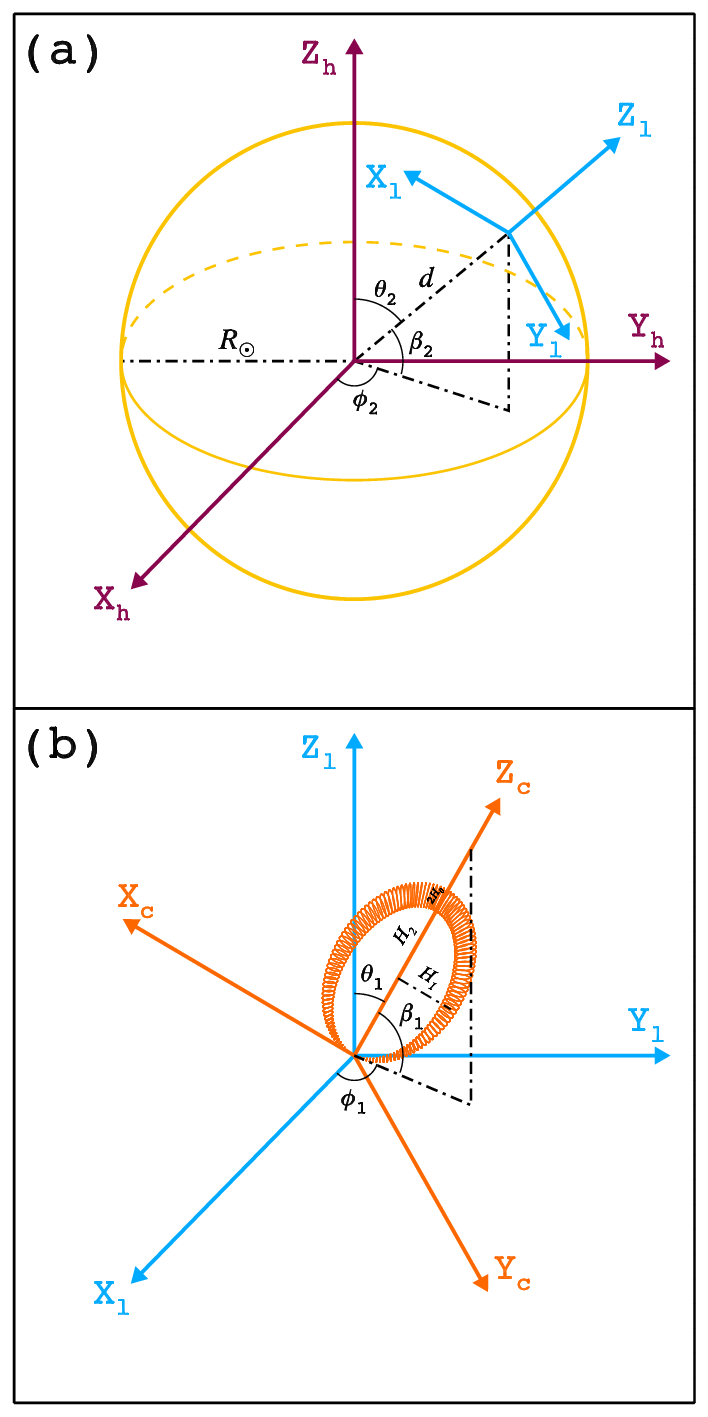}
   \centering
    \caption{Geometrical model and the transformations of three coordinate systems. See text for details.}
    \label{fig1}
\end{figure}

\section{Geometrical model} \label{model}
As shown in Fig. 1 of \citet{mo18}, the circular torus is tapered, with the cross section changing along the torus.
The cross section reaches maximum at the torus apex and shrinks on both sides toward the Sun where the cross section is zero.
In Fig. 1 of \citet{wei21a}, the torus-like structure has a global elliptic shape, with $\rho_0$ and $\rho_1$ denoting the major and minor radius of the base torus, respectively.
The cross section is also elliptic, which is characterized by an ellipticity ($\delta$).
\citet{rud24} took the inclination and handedness of flux ropes into account with eight types being considered in total.
Low-inclination flux ropes have smaller inclination angles with the latitude lines, 
while high-inclination flux ropes have larger inclination angles with the latitude lines (see their Fig. 1).
For the sake of simplicity, we adopt a global elliptic shape of the tapered torus with circular cross sections.
In Figure~\ref{fig1}, the major and minor radius of the base torus are denoted with $H_2$ and $H_1$ ($H_2 \ge H_1$).
The maximum radius of cross section at the apex is denoted with $H_0$. 
An expansion of the torus is manifested by growing values of $H_1$, $H_2$, and $H_0$.
The source region of an eruption is characterized by a longitude ($\phi_2$) and a colatitude ($\theta_{2}=90\degr-\beta_{2}$), where $\beta_2$ denotes the latitude.
The thinnest point is no longer located at the Sun center, but at a distance ($d$) from the Sun center ($0 \le d \le R_{\sun}$).
Like in the revised cone model \citep{zqm21} and revised GCS model \citep{zqm23}, a heliocentric coordinate system (HCS; $X_h$, $Y_h$, $Z_h$), 
a core coordinate system (CCS; $X_c$, $Y_c$, $Z_c$), and a local coordinate system (LCS; $X_l$, $Y_l$, $Z_l$) are defined first.
In HCS, $X_h$ points to Earth, $Z_h$ points to north, and $Y_h$-$Z_h$ defines the POS \citep{mich06}.
This is consistent with the Heliocentric Earth Ecliptic \citep[HEE;][]{tho06} coordinate system.
The transform between HCS and LCS is expressed as:

\begin{equation} \label{eqn-1}
\left(
\begin{array}{c}
x_h  \\
y_h  \\
z_h \\
\end{array}
\right)
=M_2
\left(
\begin{array}{c}
x_l  \\
y_l  \\
z_l \\
\end{array}
\right)
+
\left(
\begin{array}{c}
d\sin{\theta_2}\cos{\phi_2} \\
d\sin{\theta_2}\sin{\phi_2} \\
d\cos{\theta_2} \\
\end{array}
\right),
\end{equation}
where
\begin{equation} \label{eqn-2}
M_2=
\left(
\begin{array}{ccc}
\cos{\theta_2}\cos{\phi_2} & -\sin{\phi_2} & \cos{\phi_2}\sin{\theta_2}  \\
\cos{\theta_2}\sin{\phi_2}  &  \cos{\phi_2} & \sin{\phi_2}\sin{\theta_2} \\
-\sin{\theta_2}                    &         0          &  \cos{\theta_2} \\
\end{array}
\right).
\end{equation}

Two angles $\theta_1$ and $\phi_1$ stand for latitudinal and longitudinal deflections of the torus with respect to the radial direction of the source region. 
Hence, the transform between the LCS and CCS is realized by a matrix ($M_1$):
\begin{equation} \label{eqn-3}
\left(
\begin{array}{c}
x_l  \\
y_l  \\
z_l \\
\end{array}
\right)
=M_1
\left(
\begin{array}{c}
x_c  \\
y_c  \\
z_c \\
\end{array}
\right),
\end{equation}
where
\begin{equation} \label{eqn-4}
M_1=
\left(
\begin{array}{ccc}
\cos{\theta_1}\cos{\phi_1} & -\sin{\phi_1} & \cos{\phi_1}\sin{\theta_1} \\
\cos{\theta_1}\sin{\phi_1}  &  \cos{\phi_1} & \sin{\phi_1}\sin{\theta_1} \\
-\sin{\theta_1}                    &         0          &  \cos{\theta_1} \\
\end{array}
\right).
\end{equation}

The shape of the torus in the CCS is described as follows:
\begin{equation} \label{eqn-5}
\left \{
\begin{array}{ccc}
x_{c} & = & H_{tor}\sin{\theta} \\
y_{c} & = & (H_{1}+H_{tor}\cos{\theta})\cos{\phi} \\
z_{c} & = & (H_{2}+H_{tor}\cos{\theta})\sin{\phi}+H_{2}
\end{array}
\right .
\end{equation}
where
\begin{equation} \label{eqn-6}
H_{tor}=H_{0}\sin{(\frac{\phi+\pi/2}{2})}.
\end{equation}
The two parameters $\phi \in [0, 2\pi)$ and $\theta \in [0, 2\pi)$.
Like in the GCS and revised GCS models, $\gamma$ signifies the inclination angle with the EW direction. 
$\gamma >0$ ($\gamma <0$) indicates that the torus rotates in the clockwise (counterclockwise) direction, which is not shown in Figure~\ref{fig1}.

The appearances of a torus observed from STEREO and SolO in the external coordinate system (ECS; $X_e$, $Y_e$, $Z_e$) are obtained using the coordinate transforms 
associated with the locations of these spacecrafts \citep{zqm24}:
\begin{equation} \label{eqn-7}
\left(
\begin{array}{c}
x_{e}  \\
y_{e}  \\
z_{e} \\
\end{array}
\right)
=M_{0}
\left(
\begin{array}{c}
x_h  \\
y_h  \\
z_h \\
\end{array}
\right),
\end{equation}
where
\begin{equation} \label{eqn-8}
M_{0}=
\left(
\begin{array}{ccc}
\cos{\phi_{0}}\cos{\theta_{0}} & \sin{\phi_{0}}\cos{\theta_{0}} & -\sin{\theta_0} \\
-\sin{\phi_{0}} & \cos{\phi_{0}} & 0 \\
\cos{\phi_0}\sin{\theta_0}                  &     \sin{\phi_0}\sin{\theta_0}              & \cos{\theta_0} \\
\end{array}
\right).
\end{equation}
Here, $\phi_{0}$ and $\theta_{0}$ signify the longitudinal and latitudinal separations with the Sun-Earth connection.
In the current study, $\phi_{0}=26\fdg5$, $\theta_{0}=-2\fdg3$ for STA and $\phi_{0}=-58\fdg6$, $\theta_{0}=-2\fdg3$ for SolO, respectively.

\begin{figure*} 
   \includegraphics[width=0.6\textwidth,clip=]{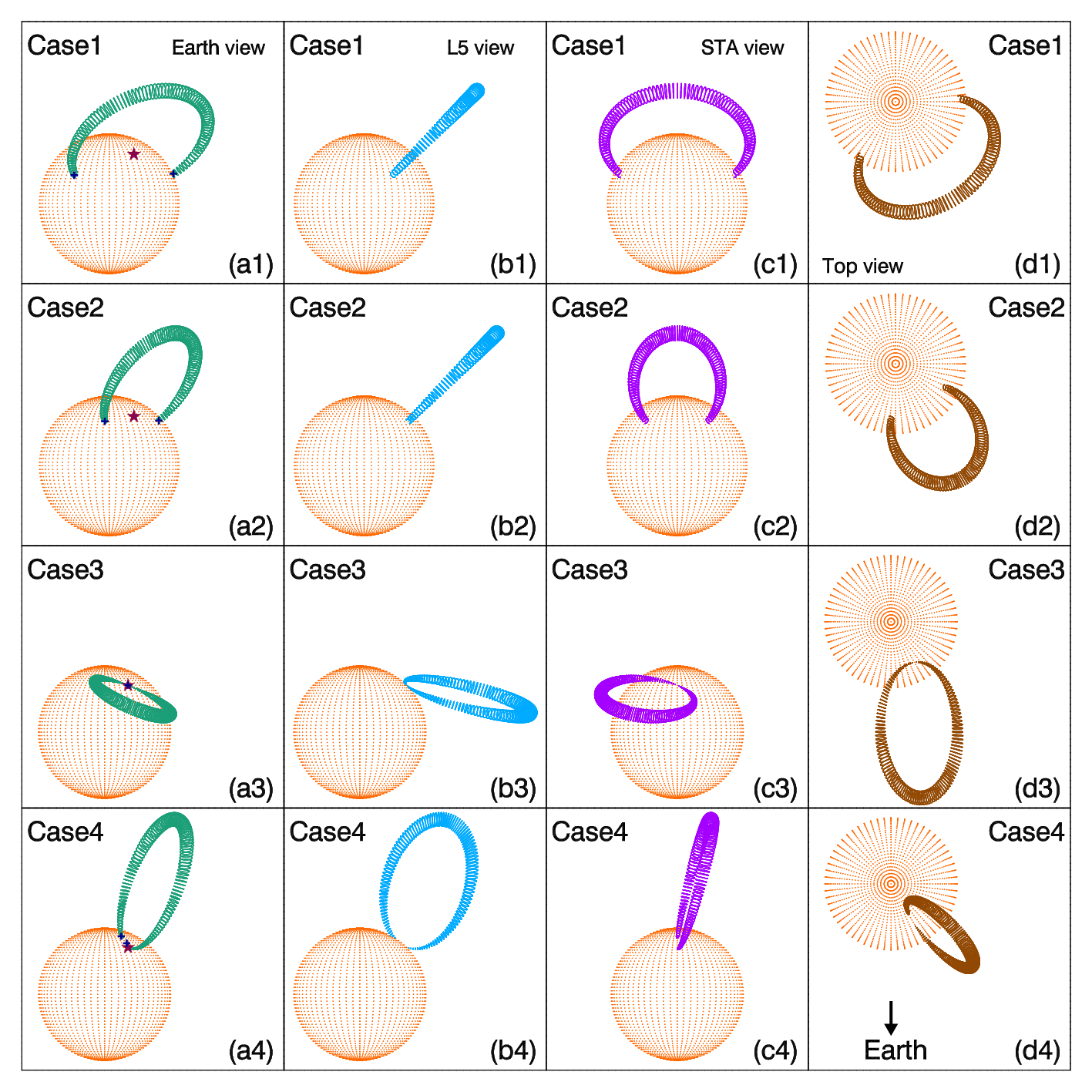}
   \centering
    \caption{The Sun and four artificial flux ropes (Case1-Case4) observed from four perspectives: 
    Earth (a1)-(a4), L5 (b1)-(b4), STA (c1)-(c4), and solar north pole (d1)-(d4).
    In the left panels, the deep red stars represent source regions of the flux ropes.
    The deep blue pluses denote footpoints of the flux ropes in the photosphere.
    In Case1 and Case2, there are no deflection or inclination. 
    The thinnest point is 0.2\,$R_{\sun}$ and 0.6\,$R_{\sun}$ from the Sun center, respectively.
    In Case3 and Case4, the thinnest point is located on the solar surface. 
    The flux rope propagates nonradially with southward and northward deflection, respectively.}
    \label{fig2}
\end{figure*}

Figure~\ref{fig2} shows the Sun and four artificial flux ropes (Case1$-$Case4) in our model as observed from four perspectives: 
Earth (a1)-(a4), Lagrange 5 (L5) point (b1)-(b4), STA (c1)-(c4), and solar north pole (d1)-(d4), 
assuming that STA has a separation angle of 30$\degr$ with the Sun-Earth line.
Parameters of the four cases are listed in Table~\ref{tab1}. 
The source region is the same with longitude and latitude being 30$\degr$ and 45$\degr$ (deep red stars in the left panels).
Consequently, the source region is just at the western limb observed from L5 point and is along the meridian observed from STA.
In other words, the observer has an edge-on view and a face-on view of the torus from L5 point and STA, respectively.
In Case1, the global shape of the torus is a circle with a radius of 725 Mm. The direction of the torus is exactly radial ($\theta_1=0$, $\phi_1=0$).
Meanwhile, the torus has no inclination with respect to EW direction ($\gamma=0$). The thinnest point has a distance of 0.2\,$R_{\sun}$ from the Sun center.
Therefore, the footpoints of the flux rope at the solar surface is far from each other (deep blue pluses in panel (a1)).

In the following three cases, the global shape of the torus is an ellipse with major and minor radii of 725 and 435 Mm, respectively.
In Case2, the torus is still radial and has no inclination. The thinnest point has a distance of 0.6\,$R_{\sun}$ from the Sun center.
The distance between the footpoints of the flux rope at the solar surface becomes much shorter (deep blue pluses in panel (a2)).
The flare ribbons as a result of flux rope eruption would be parallel to the line connecting the two footpoints.
In Case3 and Case4, the thinnest point of the torus is on the solar surface ($d=R_{\sun}$), so that the two footpoints are very close to each other,
which is similar to the revised cone and revised GCS models.
The torus shows deflections in both longitudinal and latitudinal directions.
In Case3, it is deflected eastward by 30$\degr$ and southward by 60$\degr$ (panel (c3)), suggesting that it tends to propagate equatorward \citep{zqm21,chd24,lsy25b}.
In Case4, on the contrary, it is deflected eastward by 30$\degr$ and northward by 30$\degr$ (panel (c4)), meaning that it tends to propagate poleward \citep{liu07,bi13}.
Besides, the torus has a notable inclination angle of 60$\degr$ with the EW direction, so that the torus looks almost longitudinal \citep{twl24}.
It is interesting that our model is quite flexible. 
The shape of the torus is quite close to that in the 3DCORE model when $\phi_1=0$, $\theta_1=0$, and $d=0$ \citep{mo18,wei21a}.
On the other hand, the shape looks like the revised GCS model when $d=R_{\sun}$ \citep{zqm23}.
When applying the model to 3D reconstructions of filaments or CMEs, 
we manually modify the values of those parameters and project the model onto images observed from multiple perspectives.
This process is repeated and the best set of parameters are derived when the projections match the observations satisfactorily.
In the next Section, we will apply it to an eruptive filament.

\begin{deluxetable*}{cccccccccccc}
		\digitalasset
		\tablewidth{\textwidth}
		\tablecaption{Parameters of the flux ropes in four cases.
		\label{tab1}}
		\tablecolumns{12}
		\tablenum{1}
		\tablehead{
		        \colhead{Case} &
			\colhead{$H_1$} &
			\colhead{$H_2$} &
			\colhead{$H_0$} &
			\colhead{$\phi_2$} &
			\colhead{$\theta_2$} &
			\colhead{$\phi_1$} &
			\colhead{$\theta_1$} &
			\colhead{$\gamma$} &
			\colhead{$d$} &
			\colhead{Deflection} &
			\colhead{Inclination} \\
			\colhead{ } &
			\colhead{(Mm)} &
			\colhead{(Mm)} &
			\colhead{(Mm)} &
			\colhead{(deg)} &
			\colhead{(deg)} &
			\colhead{(deg)} &
			\colhead{(deg)} &
			\colhead{(deg)} &
			\colhead{($R_{\sun}$)} &
			\colhead{} &
			\colhead{}
		}
		\startdata
                 Case1 & 725 & 725 & 72.5 & 30 & 45 &    0 & 0   &   0 & 0.2 &   no & no \\
                 Case2 & 435 & 725 & 72.5 & 30 & 45 &    0 & 0   &   0 & 0.6 &   no & no \\
                 Case3 & 435 & 725 & 72.5 & 30 & 45 & -30 & 60 &   0 & 1.0 & both & no \\
                 Case4 & 435 & 725 & 72.5 & 30 & 45 & -30 & -30 & 60 & 1.0 & both & yes \\
		\enddata
\end{deluxetable*}

\section{Filament eruption and 3D reconstructions} \label{obs}
\begin{figure*} 
	\includegraphics[width=0.5\textwidth,clip=]{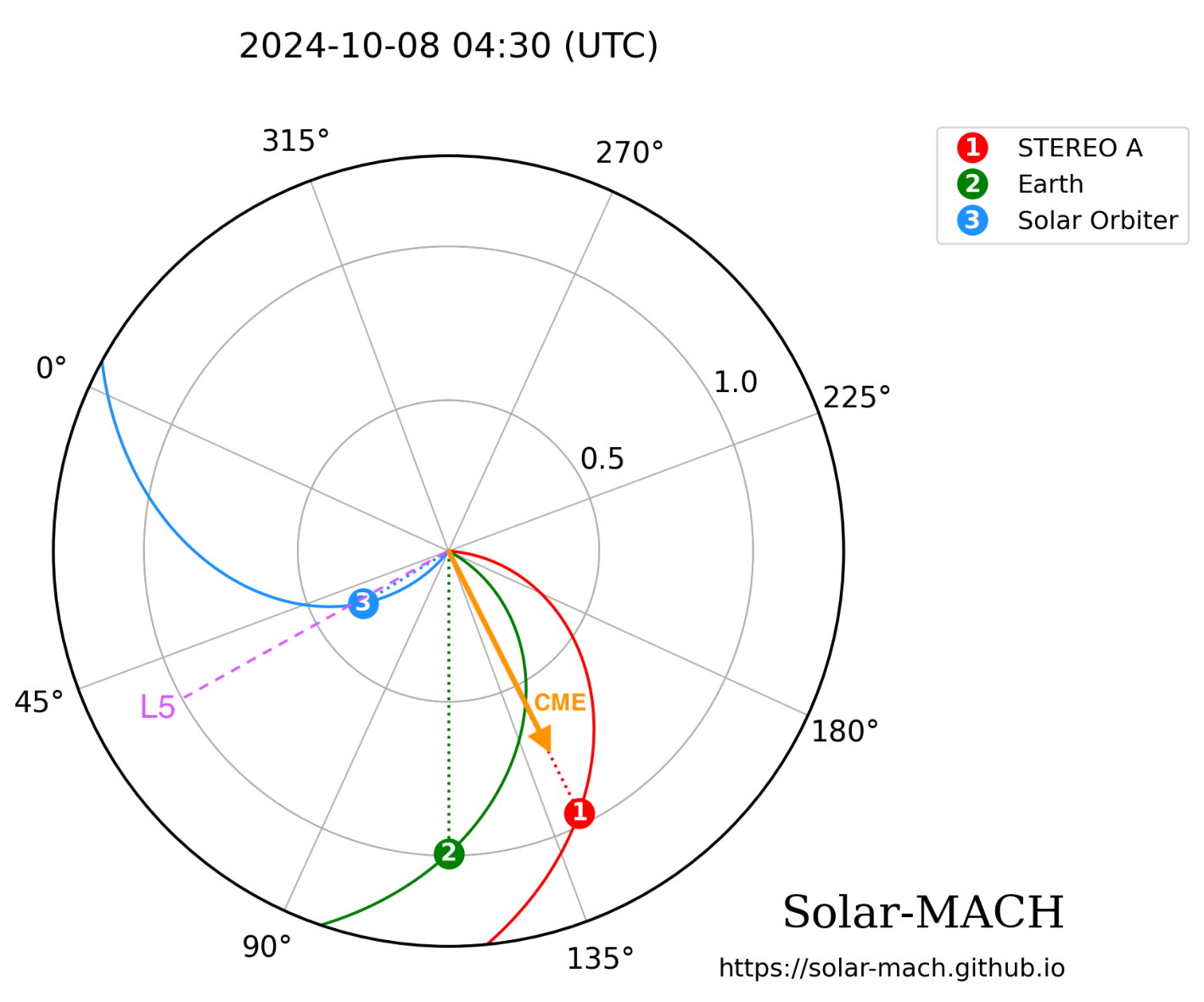}
	\centering
	\caption{The Solar-MACH plot at 04:30 UT on 2024 October 8, 
	illustrating the positions and connectivities to the Sun of Earth (green circle), STA (red circle), and SolO (blue circle).
	The position of L5 point is indicated by the pink line. The orange arrow indicates the direction of CME toward STA in the ecliptic plane.}
	\label{fig3}
\end{figure*}

On 2024 October 8, a filament erupted from AR 13847, generating three flare ribbons and a CME.
The eruption was captured by a series of spacecrafts from three perspectives.
Figure~\ref{fig3} shows the Solar-MACH plot\footnote{https://solar-mach.streamlit.app/?embedded=true} \citep{gie23} at 04:30 UT on that day.
The locations of Earth, STA, and SolO are marked with green, red, and blue circles.
STA and SolO have separation angles of 26$\fdg$5 and -58$\fdg$6 with the Sun-Earth line. SolO has almost the same longitude as L5 point (pink line).
The separation angle between STA and SolO reaches $\sim$85$\degr$, which is close to 90$\degr$. 
The heliocentric distances of these two spacecrafts are $\sim$0.960 and $\sim$0.336 AU.
Hence, the three perspectives from Earth, STA, and SolO provide an excellent opportunity for 3D reconstructions of the filament.

Apart from the Extreme-Ultraviolet Imager \citep[EUVI;][]{how08} on board STA and Extreme Ultraviolet Imager \citep[EUI;][]{ro20} on board SolO, 
the filament was also observed by the Solar Ultraviolet Imager \citep[SUVI;][]{dar22} on board the Geostationary Operational Environmental Satellite (GOES-16) spacecraft,
the Global Oscillation Network Group (GONG) in H$\alpha$ line center,
the H$\alpha$ Imaging Spectrograph \citep[HIS;][]{qiu22} on board the Chinese H$\alpha$ Solar Explorer \citep[CHASE;][]{li22}, 
and the Solar Corona Imager in UV band (SCIUV) of the Lyman-alpha (Ly$\alpha$) Solar Telescope \citep[LST;][]{feng19,li19}
on board the Advanced Space-based Solar Observatory \citep[ASO-S;][]{gan23}.
The related flare was observed by the X-Ray Telescope \citep[XRT;][]{go07} on board the Hinode \citep{ko07}, 
and the light curves in various energy bands were recorded by the Spectrometer/Telescope for Imaging X-rays \citep[STIX;][]{kru20} on board SolO.
The associated CME was observed by the C2 and C3 coronagraphs of LASCO and the COR2 coronagraph on board STA.
The full-disk line-of-sight (LOS) magnetograms of the photosphere were observed by the Helioseismic and Magnetic Imager \citep[HMI;][]{sch12} on board SDO.
The properties of instruments used in this study are summarized in Table~\ref{tab2}.

\begin{deluxetable*}{ccccccc}
		\digitalasset
		\tablewidth{\textwidth}
		\tablecaption{Properties of the instruments used in this study.
			\label{tab2}}
		\tablecolumns{4}
		\tablenum{2}
		\tablehead{
			\colhead{Instrument} &
			\colhead{Waveband} &
			\colhead{Cadence} &
			\colhead{Pixel Size}  \\
			\colhead{ } &
			\colhead{({\AA})} &
			\colhead{(s)} &
			\colhead{(arcsec)} &
		}
		\startdata
		 SDO/HMI & 6173 & 12 & 0.60 \\
		 LASCO-C2 & WL & 720 & 11.60 \\
		 LASCO-C3 & WL & 720 &  56.47 \\
                  STA/COR2 & WL & 900 & 28.24  \\
                  STA/EUVI & 304 & 600 & 1.60 \\
                  SolO/EUI & 304 & 600 & 1.50 \\
                  SolO/STIX & 4$-$25 keV & 4 & $-$ \\
                  GOES-16/SUVI & 304 & $\sim$120 & 2.50 \\
                  GONG & 6562.8 & 60 &  1.07  \\
                  CHASE/HIS & 6562.8 & 60  & 1.04  \\
                  ASO-S/SCIUV & 1216 & 60  & 2.40  \\
                  Hinode/XRT & Be\_med & 1500 & 4.11 \\
		\enddata
\end{deluxetable*}

\begin{figure} 
\includegraphics[width=0.5\textwidth]{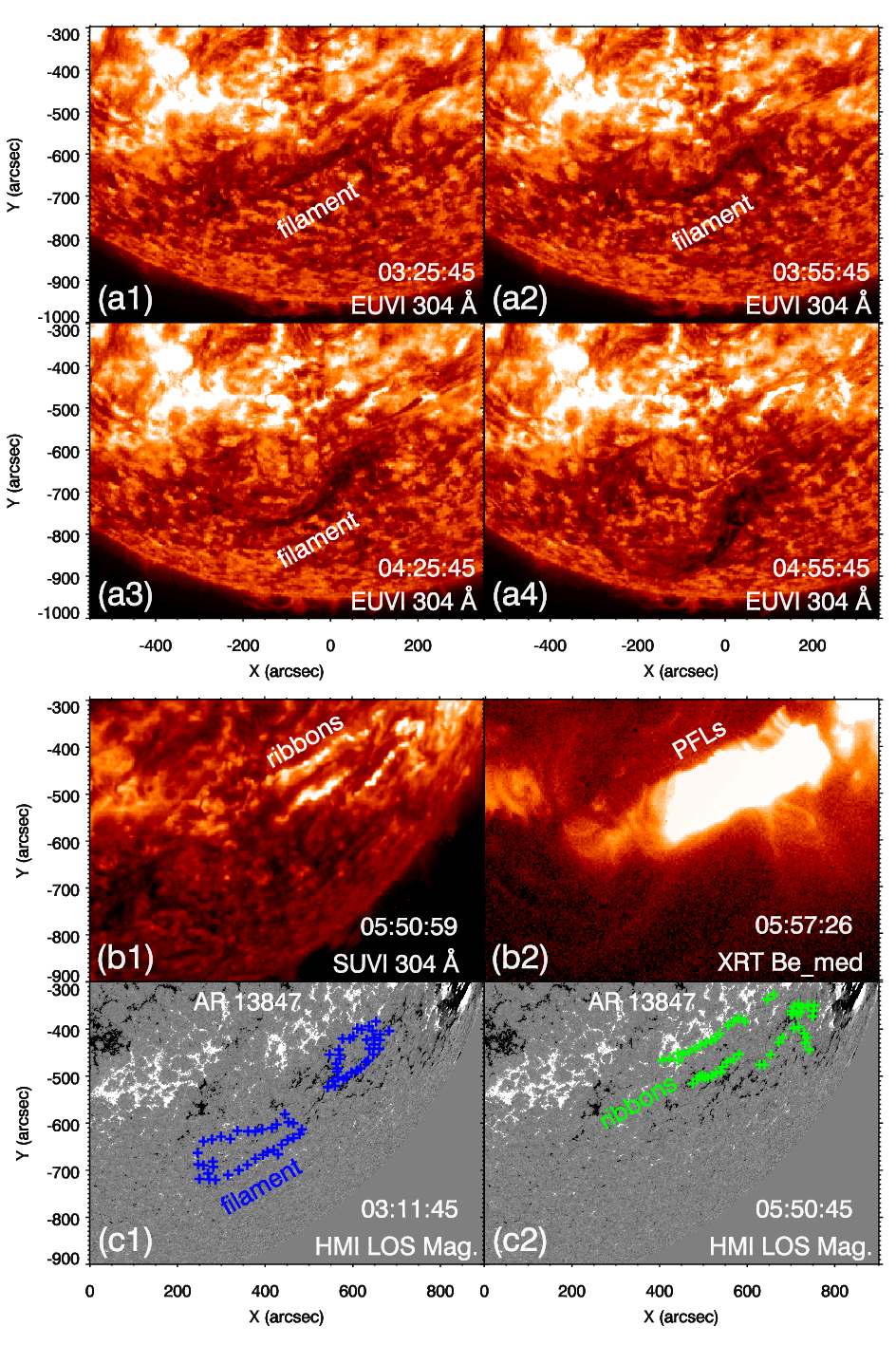}
\centering
\caption{(a1)-(a4) STA/EUVI 304 {\AA} images showing the filament eruption during 03:25$-$04:55 UT.
              (b1) SUVI 304 {\AA} image at 05:50:59 UT showing the flare ribbons.
              (b2) XRT Be\_med image at 05:57:26 UT showing the bright post-flare loops connecting the ribbons.
              (c1)-(c2) HMI LOS magnetograms of the photosphere at 03:11:45 UT and 05:50:45 UT.
              In panel (c1), outlines of the filament observed by GONG at 03:11:42 UT are superposed with blue pluses.
              In panel (c2), outlines of the flare ribbons observed by GONG at 05:50:42 UT are superposed with green pluses.
              An animation showing the filament eruption in STA/EUVI 304 {\AA} is available.
              It covers a duration of 130 minutes from 03:25 UT to 05:35 UT on 2024 October 8. 
              The entire animation runs for $\sim$1 s.
              (An animation of this figure (a1)-(a4) is available in the online article.)}
\label{fig4}
\end{figure}

In Figure~\ref{fig4}, the top four panels (a1)-(a4) are STA/EUVI 304 {\AA} images, 
showing the filament eruption in the southward direction during 03:25$-$04:55 UT (see also the online animation).
Panel (b1) shows three flare ribbons beneath the erupting filament observed by GOES-16/SUVI 304 {\AA} image 
when the brightness reaches maximum at $\sim$05:51 UT. 
The ribbons, extending in the southeast-northwest direction, are also obvious in STA/EUVI 304 {\AA}, SDO/AIA 304 {\AA}, and CHASE/HIS H$\alpha$ images.
Panel (b2) shows the bright post-flare loops (PFLs) connecting the northern and southern ribbons observed by the Hinode/XRT Be\_med filter at 05:57:26.
The Be\_med filter has a maximum response at $\log T [\mathrm{K}] \sim 7.06$ \citep{xie25}.
These PFLs are also distinct in AIA 131 {\AA} ($\log T [\mathrm{K}] \sim 7.05$) images.
The inclination angles of these ribbons with respect to the EW direction observed with different instruments are listed in Table~\ref{tab3}.
It is seen that the inclination angles are $\sim$26$\degr$ observed from STA and 30$\degr$$-$34$\degr$ observed from Earth,
which will be referable and valuable in 3D reconstructions of the filament.

Figure~\ref{fig4}(c1)-(c2) show the HMI LOS magnetograms of the photosphere at 03:11:45 UT and 05:50:45 UT.
In panel (c1), outlines of the filament observed in H$\alpha$ at the beginning of eruption ($\sim$03:11 UT) are superposed with blue pluses.
In panel (c2), outlines of the flare ribbons observed in H$\alpha$ at $\sim$05:50 UT are superposed with green pluses.
It is clear that the filament is initially located close to the polarity inversion line of AR 13847, 
and the northern (southern) ribbons are distributed in the positive (negative) polarities, respectively.

\begin{deluxetable*}{cccccc}
		\digitalasset
		\tablewidth{\textwidth}
		\tablecaption{Inclination angles of flare ribbons with respect to the EW direction observed with different instruments.
		\label{tab3}}
		\tablecolumns{5}
		\tablenum{3}
		\tablehead{
		        \colhead{Instrument} &
			\colhead{CHASE} &
			\colhead{SUVI} &
			\colhead{XRT} &
			\colhead{EUVI} \\
			\colhead{Waveband} &
			\colhead{6562.8 {\AA}} &
			\colhead{304 {\AA}} &
			\colhead{Be\_med} &
			\colhead{304 {\AA}}
		}
		\startdata
                 Angle (deg) &   30$\pm$1 & 34$\pm$1 & 30$\pm$1 & 26$\pm$2 \\
		\enddata
\end{deluxetable*}

\begin{figure} 
		\includegraphics[width=0.55\textwidth,clip=]{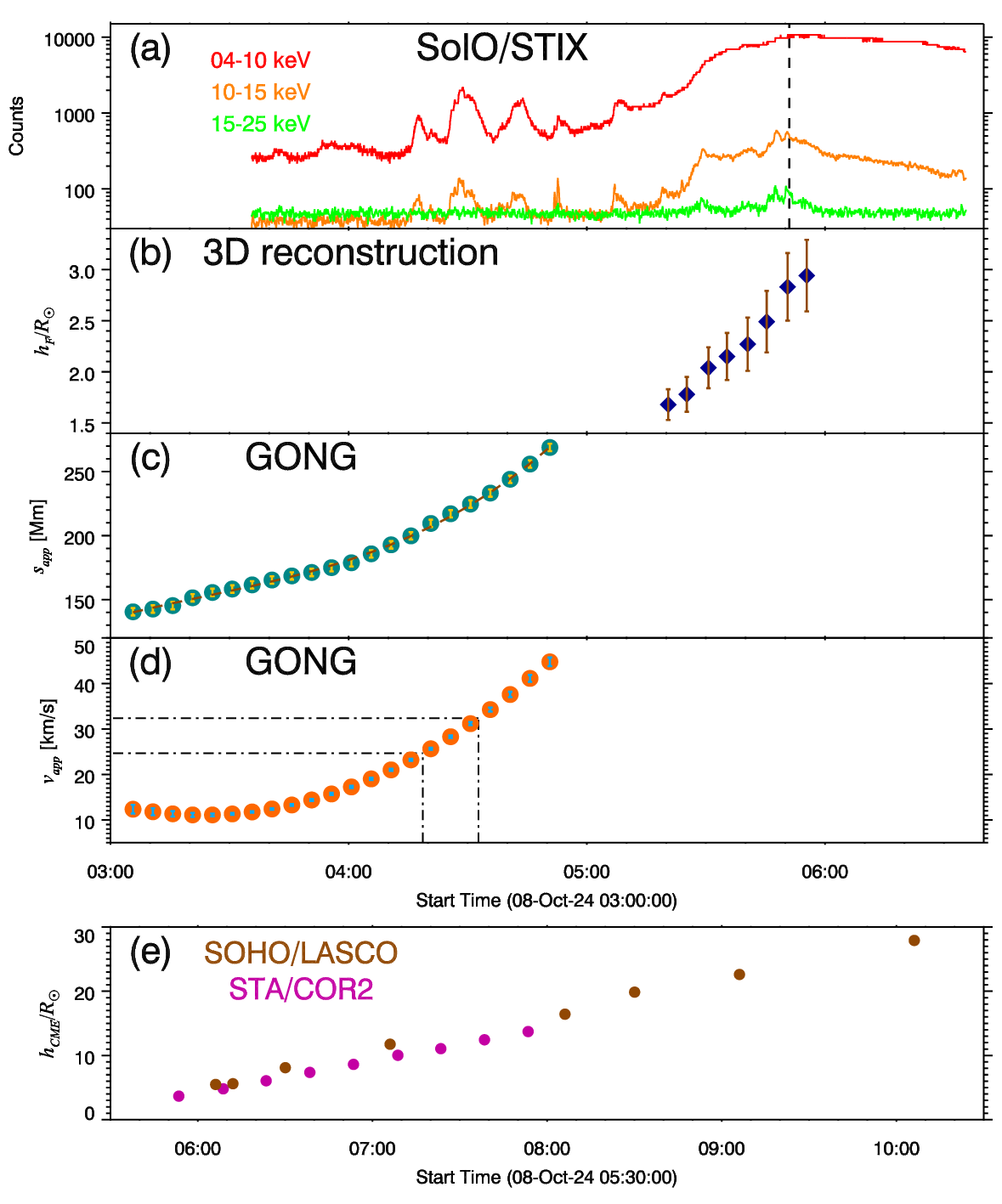}
		\centering
		\caption{(a) Light curves of the flare observed by SolO/STIX at 4$-$10 keV (red line), 10$-$15 keV (orange line), and 15$-$25 keV (green line).
		(b) Heliocentric distances of the filament leading front using 3D reconstructions.
		(c) Trajectory of the filament in the field of view (FOV) of GONG. A cubic fitting is conducted for the trajectory and drawn with a brown dashed line.
		(d) Apparent speeds ($v_{app}$) of the filament in the FOV of GONG. 
		The values at 04:18:18 UT and 04:32:30 UT are marked with dash-dotted lines.
		(e) Height-time plots of the CME leading edge in the FOVs of SOHO/LASCO (brown dots) and STA/COR2 (magenta dots).}
		\label{fig5}
	\end{figure}

In Figure~\ref{fig5}(a), the red, orange, and green lines signify the SolO/STIX light curves of the flare at 4$-$10, 10$-$15, and 15$-$25 keV during 03:35$-$06:35 UT.
Note that the GOES light curve in 1$-$8 {\AA} during this time range is unavailable. Hence, the GOES class of the flare is unknown.
The vertical black line denotes the flare peak time at $\sim$05:51 UT.
During the impulsive phase, five small peaks could be notably identified at 4$-$10 and 10$-$15 keV, implying significant energy releases of the flare.

\begin{figure} 
\includegraphics[width=0.55\textwidth]{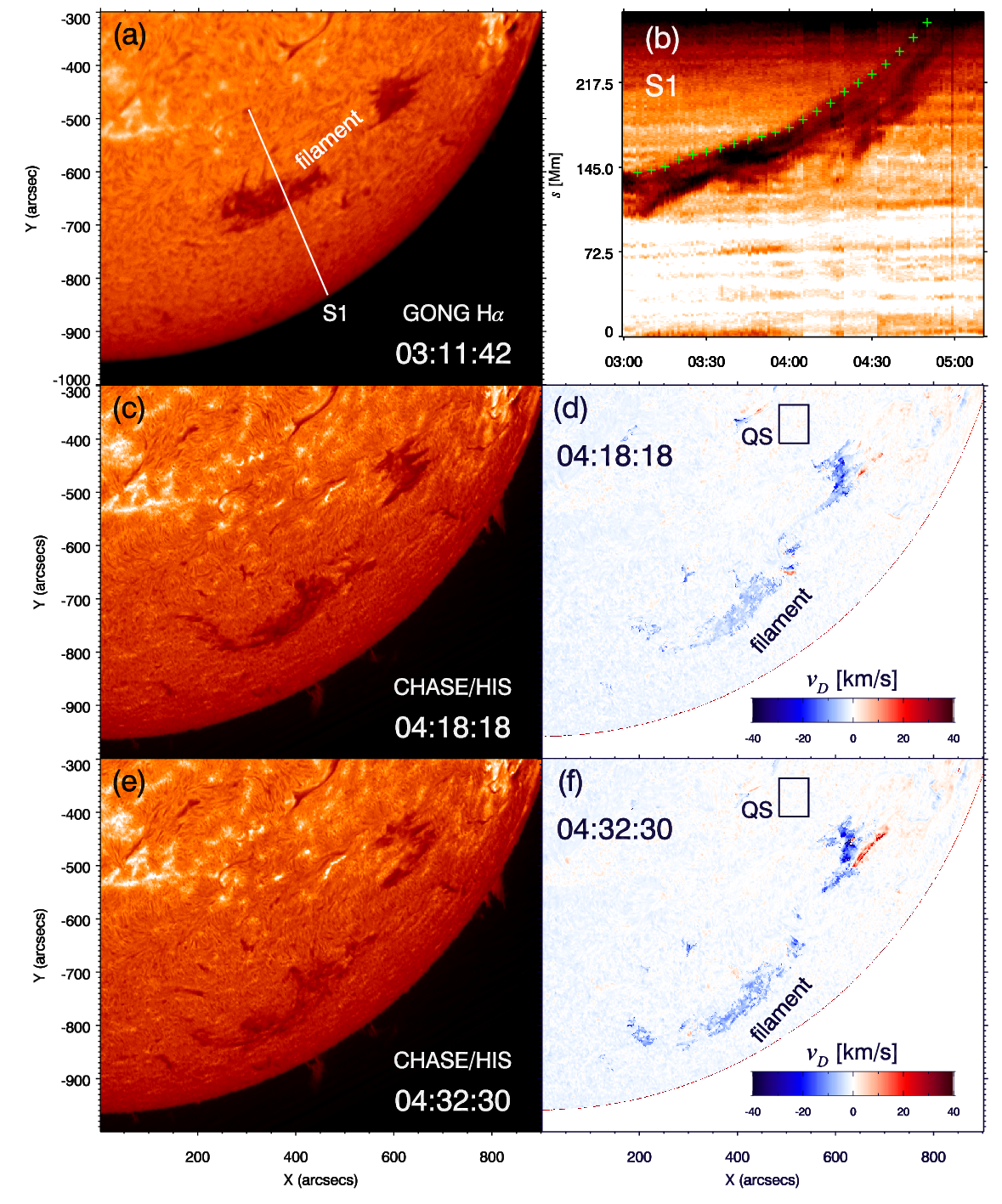}
\centering
\caption{(a) GONG H$\alpha$ image at 03:11:42 UT. The slice S1 is along the direction of filament eruption.
             (b) Time-slice diagram of S1 in H$\alpha$ during 03:00$-$05:10 UT. 
             The green pluses represent the trajectory of the filament leading front along S1.
             (c)-(d) Intensity map and Dopplergram observed by CHASE/HIS at 04:18:18 UT.
             The black box denotes the quiet region (QS).
             (e)-(f) The same maps taken at 04:32:30 UT.}
\label{fig6}
\end{figure}

In Figure~\ref{fig6}, panel (a) shows the GONG H$\alpha$ image at 03:11:42 UT. The filament is composed of two parts.
To investigate its evolution in the POS, we chose a slice S1 along the direction of eruption with a length of 278.4 Mm.
The time-slice diagram of S1 in H$\alpha$ is shown in panel (b). 
It is seen that the dark filament rises up gradually from $\sim$03:10 UT and accelerates later until $\sim$05:00 UT when the filament moves out of the disk.
Trajectory of the filament in the POS is plotted with green pluses in Figure~\ref{fig6}(b) and green circles in Figure~\ref{fig5}(c).
A cubic fitting is applied to the trajectory and superposed with a brown dashed line in Figure~\ref{fig5}(c).
The associated velocities ($v_{app}$) of the filament in the POS are drawn with orange circles in Figure~\ref{fig5}(d).
The values of $v_{app}$ fall in the range of 11$-$45 km\,s$^{-1}$ during 03:10$-$04:50 UT.

In Figure~\ref{fig6}, panels (c) and (e) show the filament observed by CHASE/HIS at 04:18:18 UT and 04:32:30 UT.
The related Dopplergrams are displayed in panels (d) and (f), respectively.
The quiet region (QR) within the black box (60$\arcsec \times 72\arcsec$) is used to calculate the reference spectrum (wavelength) of H$\alpha$ line \citep{qiu24}.
The filament is dominated by blueshift \citep{chc25}.
As the filament lifts off and accelerates, it becomes loose and the brightness increases.
The average blue-shift velocity ($v_D$) increases from $\sim$8 km\,s$^{-1}$ at 04:18 UT to $\sim$12 km\,s$^{-1}$ at 04:32:30 UT.
In Figure~\ref{fig5}(d), the apparent speeds of the filament are $\sim$24.6 and $\sim$32.4 km\,s$^{-1}$ at 04:18 UT and 04:32 UT (dash-dotted lines), respectively.
The associated true speeds ($v_{true}=\sqrt{v_{app}^2+v_{D}^2}$) are calculated to be $\sim$25.9 and $\sim$34.5 km\,s$^{-1}$.
The included angles between $v_{true}$ and LOS are $\sim$72$\degr$ and $\sim$70$\degr$ at two moments,
indicating that the direction of filament eruption is significantly deflected southward (poleward).

\begin{figure} 
		\includegraphics[width=0.90\textwidth,clip=]{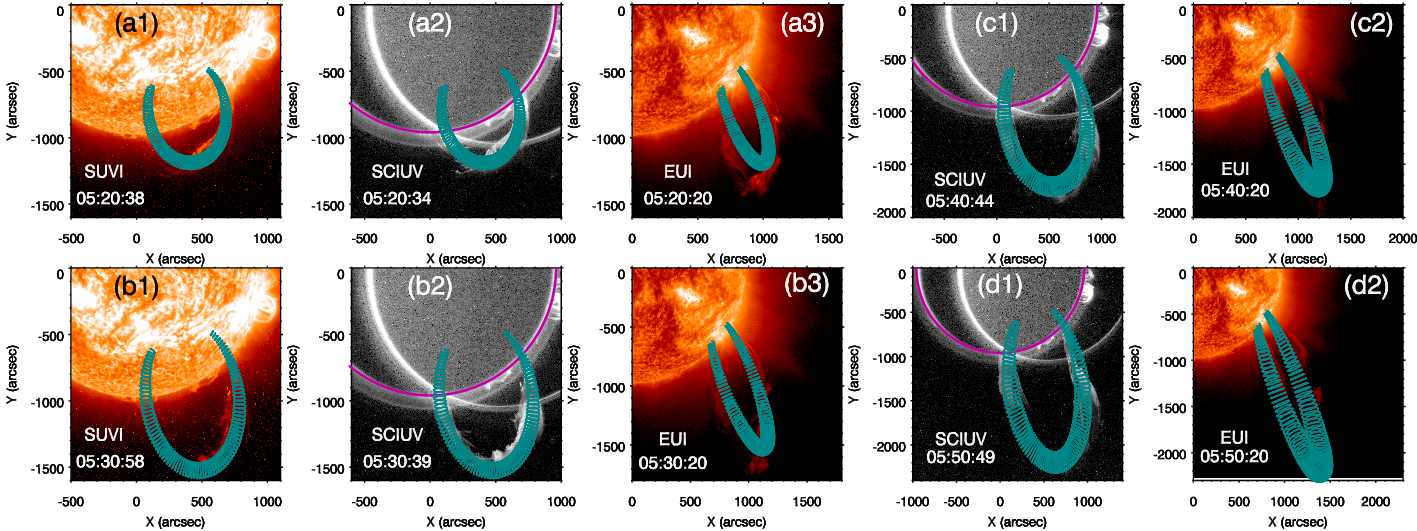}
		\centering
		\caption{The eruptive filament observed by SUVI 304 {\AA}, SCIUV Ly$\alpha$, and EUI 304 {\AA} 
		at 05:20 UT (a1)-(a3), 05:30 UT (b1)-(b3), 05:40 UT (c1)-(c2), and 05:50 UT (d1)-(d2).
		The reconstructed tori (flux ropes) are superposed with turquoise dots.}
		\label{fig7}
	\end{figure}

After 05:00 UT, the leading front of the filament leaves the disk and shows up as a huge prominence viewed from three perspectives (Earth, STA, and SolO),
which facilitate 3D reconstructions of the filament using the geometrical model in Section~\ref{model}.
It is noted that simultaneous observations from two perspectives are adequate to carry out reconstructions, since the time cadences of these instruments are not equal.
Besides, the images observed with STA and SolO are rescaled by using the standard SSW routine \texttt{scale\_map.pro}, 
so that the heliocentric distances of STA and SolO are equal to that of Earth.
The large FOVs of ASO-S/SCIUV and SolO/EUI allow for tracking the filament continuously for 35 minutes 
until 05:55 UT when the CME leading edge has come out in the FOV of STA/COR2.
In Figure~\ref{fig7}, the filament observed by SUVI, SCIUV, and EUI at 05:20 UT are displayed in panels (a1), (a2), and (a3). In panel (a2), the purple arc outlines the solar limb.
It is noticed that the bright circular feature seen in SCIUV coronagraph is due to an unexpected flaw of the instrument, 
which causes contamination of the real Ly$\alpha$ images. The false feature should be ignored.
Projections of the reconstructed torus onto these images are overlaid with turquoise dots.
It is obvious that the model fits well with the filament, not only at the leading front, but also at the legs.
The fitted parameters are listed in Table~\ref{tab4}.
At 05:20 UT, the values of $H_1$, $H_2$, and $H_0$ are 217.5$\pm$2.8, 304.5$\pm$3.2, and 36.3$\pm$1.2 Mm.
The source region of the filament is located at W30S30. 
The filament is slightly deflected eastward by $\sim$10$\degr$ and notably deflected southward by $\sim$40$\degr$ at the same time.
The value of $\gamma$ is -30$\degr$, which is nicely in agreement with the inclination angles of flare ribbons (Figure~\ref{fig4} and Table~\ref{tab3}).
The value of $d$ is $\sim$0.85\,$R_{\sun}$ and the heliocentric distance of filament leading front ($h_F$) reaches $\sim$1.68\,$R_{\sun}$.
For each parameter of $H_1$, $H_2$, $H_0$, and $\phi_{2}$ we manually adjust the values for 10 times 
and project the modeled flux ropes onto the SUVI, SCIUV, EUI, and EUVI images when the fittings are still satisfactory.
The standard deviations are considered as errors of the fitted parameters.

In Figure~\ref{fig7}, panels (b1)-(b3) show the filament observed by SUVI, SCIUV, and EUI at 05:30 UT, with the reconstructed torus superposed with turquoise dots.
As the CME expands during eruption, the values of $H_1$, $H_2$, and $H_0$ increase to 261.0$\pm$3.7, 435.0$\pm$4.2, and 43.5$\pm$1.5 Mm.
The longitude increases slightly by $\sim$2$\degr$, while the latitude is invariant. 
The values of $\theta_2$, $\theta_1$, $\phi_1$, $\gamma$, and $d$ do not change, 
implying that the filament evolves in a self-similar manner and propagates almost in the same direction.
After 05:30 UT, the filament leading front escapes from the FOV of SUVI and is visible only in the images of SCIUV (panels (c1), (d1)) and EUI (panels (c2), (d2)).
It is evident that the model still fits the filament well 30 minutes later.

\begin{figure} 
   \includegraphics[width=0.90\textwidth,clip=]{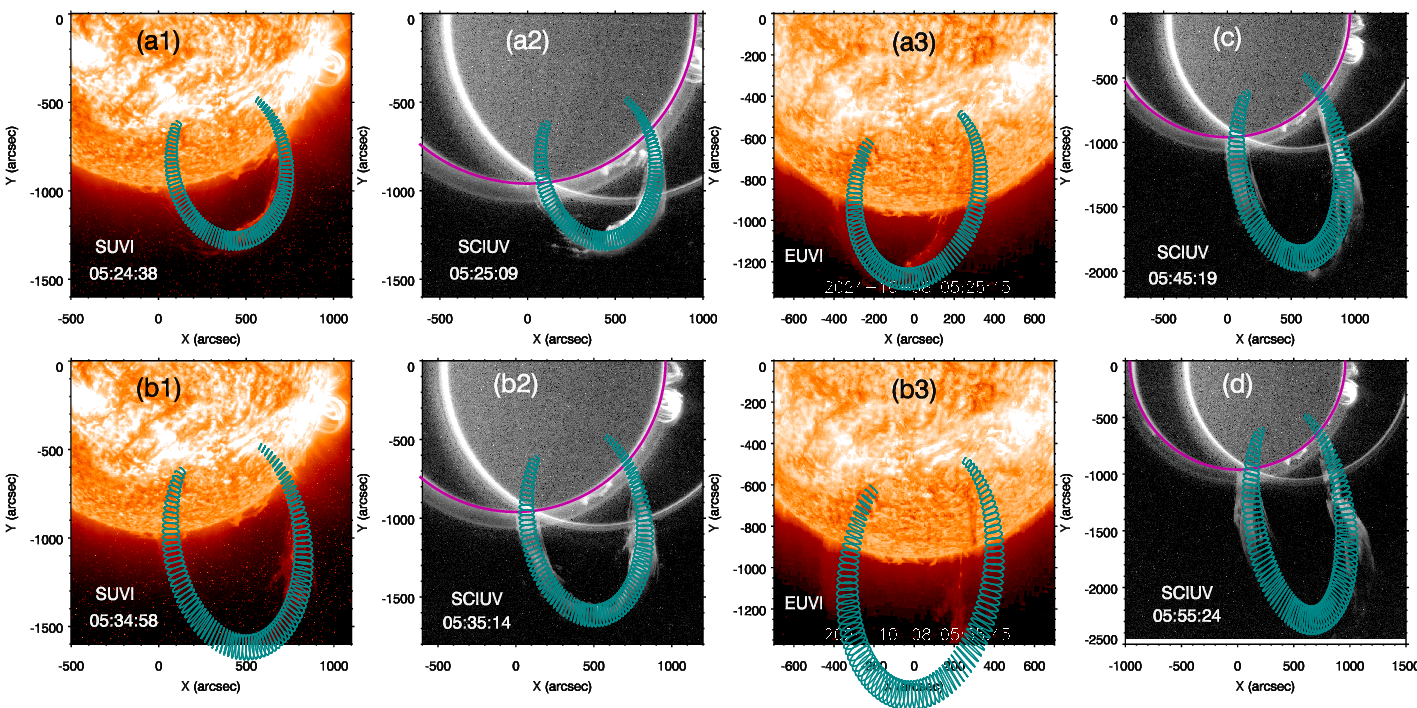}
   \centering
   \caption{The eruptive filament observed by SUVI 304 {\AA}, SCIUV Ly$\alpha$, and EUVI 304 {\AA} 
		at 05:25 UT (a1)-(a3), 05:35 UT (b1)-(b3), 05:45 UT (c), and 05:55 UT (d).
		The reconstructed tori (flux ropes) are superposed with turquoise dots.}
   \label{fig8}
\end{figure}

In Figure~\ref{fig8}, panels (a1)-(a3) and (b1)-(b3) show the filament observed by SUVI, SCIUV, and EUVI at 05:25 UT and 05:35 UT, respectively.
The reconstructed tori are superposed with turquoise dots. It is found that the tori fit the filament well viewed from Earth.
In EUVI images, the tori fit well with the leading front of the filament.
There is slight discrepancy between the observation and model at the legs, which is probably due to that the filament is not exactly coplanar.

\begin{figure} 
		\includegraphics[width=0.75\textwidth,clip=]{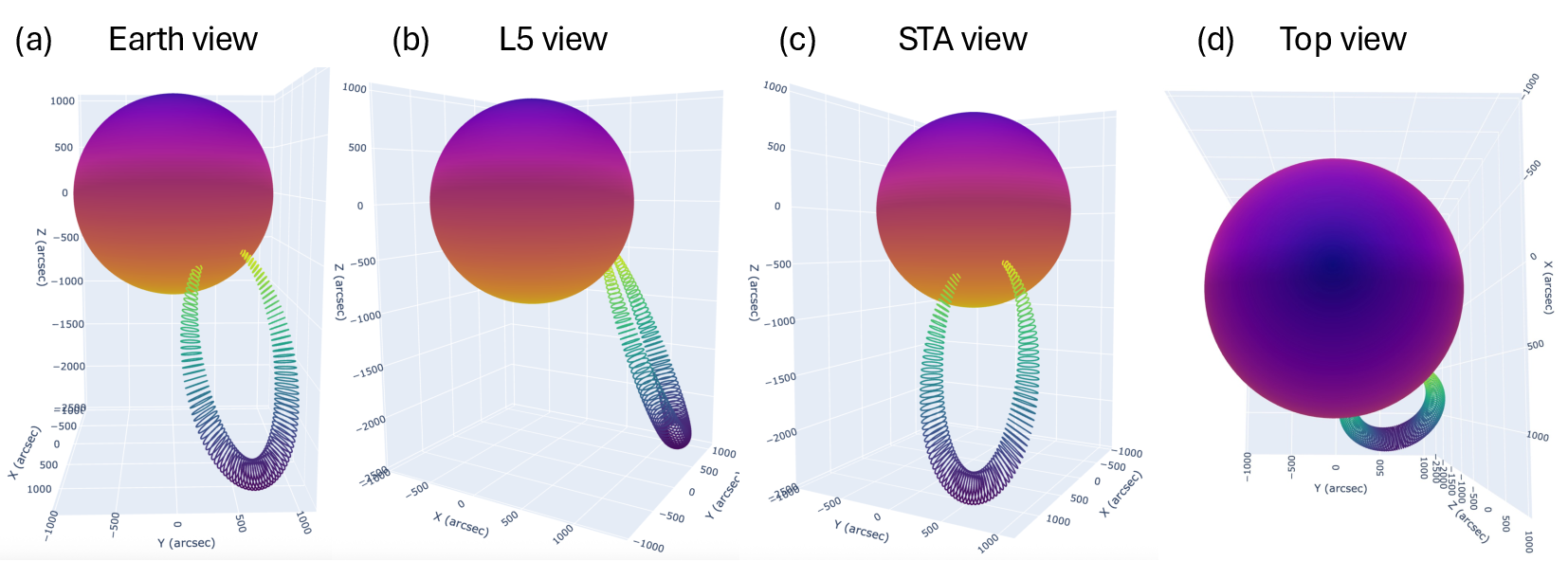}
		\centering
		\caption{The 3D visualization of the Sun and reconstructed torus at 05:55 UT viewed from Earth (a), L5 point (b), STA (c), and solar north pole (d).}
		\label{fig9}
	\end{figure}

After 05:35 UT, the filament has escaped from the FOV of STA/EUVI. As the filament moves on, only observations from SCIUV are available at 05:45 UT and 05:55 UT.
Therefore, we fix the direction ($\phi_1$, $\theta_1$, $\gamma$), colatitude ($\theta_2$), and $d$, 
and just modify the values of $H_1$, $H_2$, $H_0$, and $\phi_2$ at these two moments.
Panels (c) and (d) show the SCIUV images and projections of the tori, indicating that the fitting is acceptable until 05:55 UT. 
In Figure~\ref{fig9}, the Sun and reconstructed torus at 05:55 UT as viewed from Earth (a), L5 point (b), STA (c), and solar north pole (d)
are visualized using the Plotly Python Graphing Library\footnote{https://plotly.com/python/}.

\begin{figure}
\includegraphics[width=0.50\textwidth,clip=]{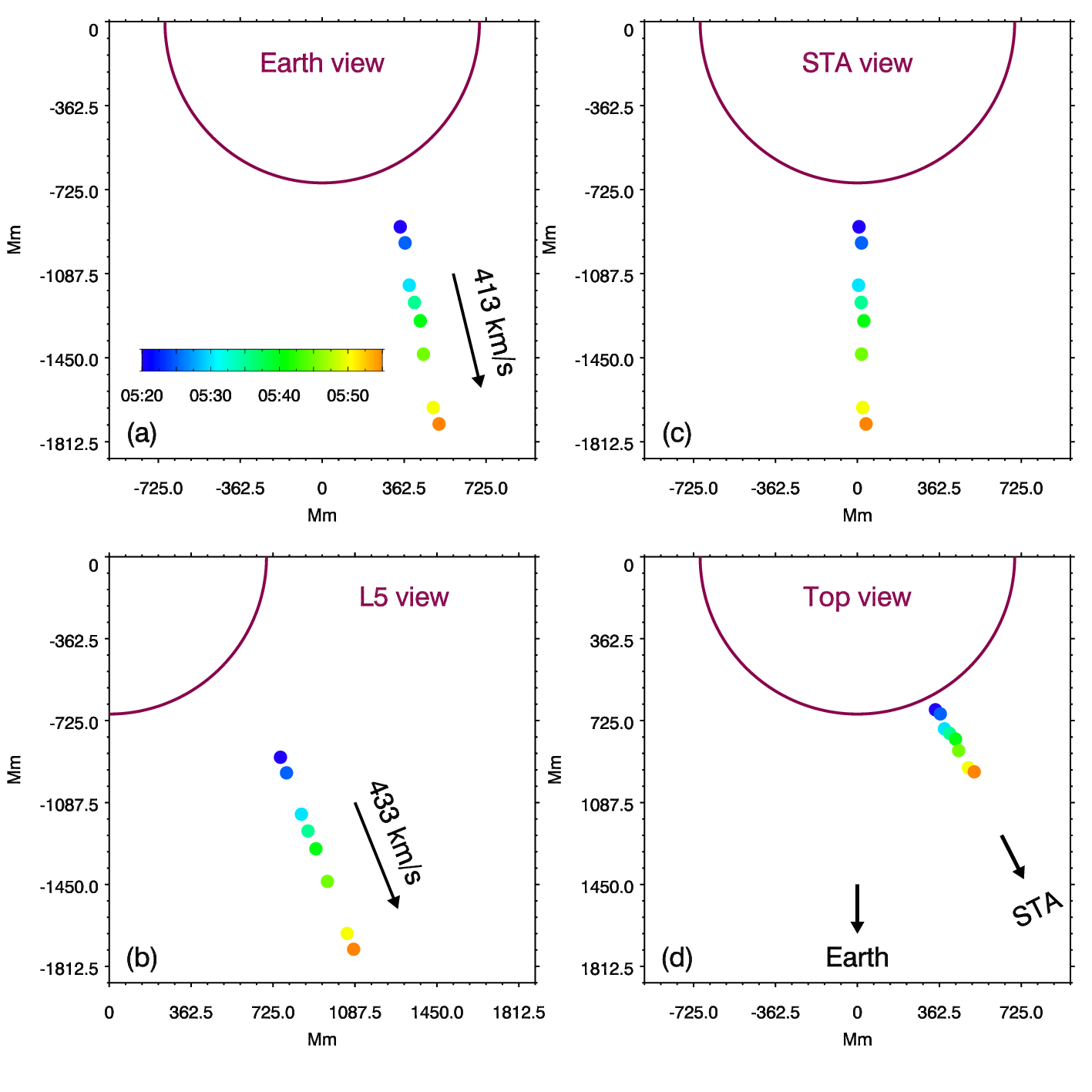}
\centering
\caption{Trajectory of the filament leading front viewed from Earth (a), L5 point (b), STA (c), and solar north pole (d) during 05:20$-$05:55 UT.
              Apparent velocities of filament are labeled in the left panels. 
              In panel (d), the trajectory shows that the filament heads for STA instead of Earth in the ecliptic plane.}
\label{fig10}
\end{figure}

The parameters of 3D reconstructions from 05:20 UT to 05:55 UT with an interval of 5 minutes are listed in Table~\ref{tab4}. 
In Figure~\ref{fig5}(b), the deep blue diamonds signify the heliocentric distances of the leading front ($h_F$), 
which increase from $\sim$1.68 to $\sim$2.94\,$R_{\sun}$ within 35 minutes.
In Figure~\ref{fig10}, the colored dots represent time evolution of the filament leading front viewed from Earth (a), L5 point (b), STA (c), and solar north pole (d).
In panel (a), the apparent speed of the filament in the POS is $\sim$413 km\,s$^{-1}$.
Since the projection effect of the filament eruption almost vanishes from the perspective of L5 point, the speed of filament ($\sim$433 km\,s$^{-1}$) in panel (b) 
represents its true speed during 05:20$-$05:55 UT. 
Accordingly, the included angle between the true and LOS speed is $\sim$72$\fdg$5, 
which is consistent with the value at 04:18 UT using spectral observations (Figure~\ref{fig6}).
In other words, the true direction of the eruptive filament is the same using two independent methods, i.e., 3D reconstruction and spectral observation,
which confirms the reliability of our geometrical model.
In panel (d), the trajectory shows that the eruptive filament and associated CME accurately heads for STA in the ecliptic plane (see orange arrow in Figure~\ref{fig3}).

\begin{deluxetable*}{cccccccccccc}
		\digitalasset
		\tablewidth{\textwidth}
		\tablecaption{Parameters of 3D reconstruction of the filament at eight moments with a cadence of 5 minutes.
		\label{tab4}}
		\tablecolumns{12}
		\tablenum{4}
		\tablehead{
		        \colhead{Time} &
			\colhead{$H_1$} &
			\colhead{$H_2$} &
			\colhead{$H_0$} &
			\colhead{$\phi_2$} &
			\colhead{$\theta_2$} &
			\colhead{$\beta_2$} &
			\colhead{$\phi_1$} &
			\colhead{$\theta_1$} &
			\colhead{$\gamma$} &
			\colhead{$d$} &
			\colhead{$h_{F}$} \\
			\colhead{(UT)} &
			\colhead{(Mm)} &
			\colhead{(Mm)} &
			\colhead{(Mm)} &
			\colhead{(deg)} &
			\colhead{(deg)} &
			\colhead{(deg)} &
			\colhead{(deg)} &
			\colhead{(deg)} &
			\colhead{(deg)} &
			\colhead{($R_{\sun}$)} &
			\colhead{($R_{\sun}$)}
		}
		\startdata
                 05:20:31 &  217.5$\pm$2.8 & 304.5$\pm$3.2 & 36.3$\pm$1.2 & 30$\pm$1 & 120 & -30 & -10 & 40 & -30 & 0.85 & 1.68$\pm$0.15 \\
                 05:25:11 &   239.3$\pm$3.5 & 340.8$\pm$3.4 & 37.7$\pm$1.3 & 31$\pm$1 & 120 & -30 & -10 & 40 & -30 & 0.85 & 1.78$\pm$0.17 \\
                 05:30:39 &   261.0$\pm$3.7 & 435.0$\pm$4.2 & 43.5$\pm$1.5 & 32$\pm$1 & 120 & -30 & -10 & 40 & -30 & 0.85 & 2.04$\pm$0.20 \\
                 05:35:19 &   275.5$\pm$4.1 & 471.3$\pm$4.3 & 50.8$\pm$1.8 & 33$\pm$1 & 120 & -30 & -10 & 40 & -30 & 0.85 & 2.15$\pm$0.23 \\ 
                 05:40:32 &   290.0$\pm$4.3 & 507.5$\pm$4.7 & 65.3$\pm$2.3 & 34$\pm$1 & 120 & -30 & -10 &  40 & -30 & 0.85 & 2.27$\pm$0.26 \\     
                 05:45:19 &   304.5$\pm$4.6 & 580.0$\pm$4.9 & 72.5$\pm$2.5 & 35$\pm$1 & 120 & -30 & -10 & 40 & -30 & 0.85 & 2.49$\pm$0.30 \\  
                 05:50:35 &   304.5$\pm$4.6 & 696.0$\pm$5.5 & 87.0$\pm$2.7 & 36$\pm$1 & 120 & -30 & -10 &  40 & -30 & 0.85 & 2.83$\pm$0.33 \\ 
                 05:55:24 &   304.5$\pm$4.6 & 732.3$\pm$5.7 & 90.6$\pm$3.0 & 37$\pm$1 & 120 & -30 & -10 & 40 & -30 & 0.85 & 2.94$\pm$0.35 \\
		\enddata
\end{deluxetable*}

\begin{figure} 
		\includegraphics[width=0.60\textwidth,clip=]{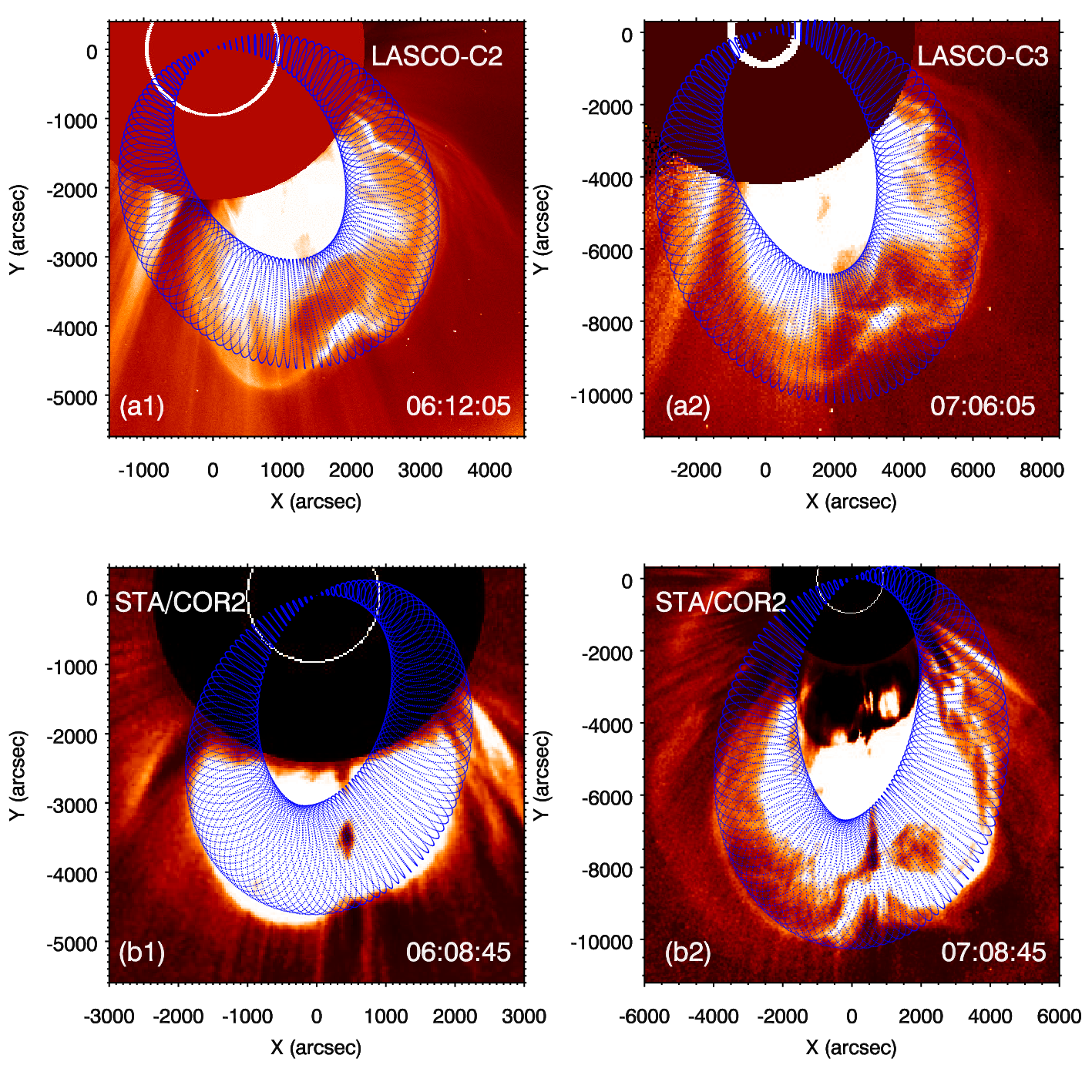}
		\centering
		\caption{The related CME observed by SOHO/LASCO (upper panels) and STA/COR2 (lower panels) around 06:10 UT and 07:07 UT.
		The blue dots represent 3D reconstruction of the CME front with 3DCORE model.}
		\label{fig11}
	\end{figure}

In Figure~\ref{fig11}, the upper panels show the CME driven by the filament eruption observed by LASCO-C2 at 06:12:05 UT (a1) and LASCO-C3 at 07:06:05 UT (a2).
The partial-halo CME is recorded by the CDAW CME catalog\footnote{https://cdaw.gsfc.nasa.gov/CME\_list/} \citep{gop09}.
The lower panels show the CME observed by STA/COR2 at 06:08:45 UT (b1) and 07:08:45 UT (b2).
Angular widths of the CME are $\sim$90$\degr$ and $\sim$100$\degr$ in LASCO and COR2 images.
Like the event on 2007 November 14, the CME leading edge distorts into a concave structure \citep{sav10,lie11}.
The directions of CME in the FOVs of LASCO and STA are generally in agreement with those of filament in the FOVs of SUVI and EUVI.
In Figure~\ref{fig5}(e), the brown and magenta dots represent the heliocentric distances of the CME leading edge ($h_{CME}$) observed by LASCO and COR2, respectively.
The value of $h_{CME}$ grows from $\sim$3.6\,$R_{\sun}$ at 05:53 UT to $\sim$27.9\,$R_{\sun}$ at 10:06 UT.
Linear fittings of the height-time plots during 05:53$-$10:06 UT result in apparent velocities of $\sim$1092 and $\sim$978 km\,s$^{-1}$.
Assuming that the true direction of CME leading edge is consistent with that of filament,
the true velocities of CME are estimated to be $\sim$1145 and $\sim$1046 km\,s$^{-1}$ based on 3D reconstructions of the filament during 05:20$-$05:55 UT.
The marginal discrepancy between the estimated true speeds using observations from two perspectives 
is presumably due to that the CME and filament are not exactly homodromous.
In Figure~\ref{fig11}, the blue dots represent 3D reconstruction of the CME front with 3DCORE model, 
in which the thinnest point of a flux rope is located at the Sun center \citep{mo18}. It is clear that the leading edges of flux ropes are generally cospatial with the CME fronts.
The associated parameters are listed in Table~\ref{tab5}. The value of $\phi_2$ ($\sim$32$\degr$) is lower than most of the values for the filament during 05:20$-$05:55 UT, 
indicating that the CME deflects eastward from the source region. 
Besides, the values of $\beta_2$ decreases from -50$\degr$ to -60$\degr$, suggesting a southward deflection of the CME.
The heliocentric distance of the CME front increases from $\sim$5.94\,$R_{\sun}$ to $\sim$11.98\,$R_{\sun}$.
\citet{wang26} analyzed a CME interaction event that gave rise to a major geomagnetic storm on 2024 October 10-11.
The two interacting CMEs result from sympathetic filament eruptions on 2024 October 9.
They also checked the CME starting at 06:12 UT on October 8, which is studied in this paper.
Using Wang-Sheeley-Arge (WSA)-ENLIL simulations, they found that the CME was finally directed too far south to impact Earth,
which is in line with our results that the filament and CME deflect southward by $\sim$40$\degr$.

\begin{deluxetable*}{ccccccccc}
		\digitalasset
		\tablewidth{\textwidth}
		\tablecaption{Parameters of 3D reconstruction of the CME front around 06:10 UT and 07:07 UT.
		\label{tab5}}
		\tablecolumns{9}
		\tablenum{5}
		\tablehead{
		        \colhead{Time} &
			\colhead{$H_1$} &
			\colhead{$H_2$} &
			\colhead{$H_0$} &
			\colhead{$\phi_2$} &
			\colhead{$\theta_2$} &
			\colhead{$\beta_2$} &
			\colhead{$\gamma$} &
			\colhead{$r_{CME}$}\\
			\colhead{(UT)} &
			\colhead{(Mm)} &
			\colhead{(Mm)} &
			\colhead{(Mm)} &
			\colhead{(deg)} &
			\colhead{(deg)} &
			\colhead{(deg)} &
			\colhead{(deg)}	&
			\colhead{($R_{\sun}$)}
		}
		\startdata
                 06:10 & 1196.25 & 1776.25 & 580.00 & 32 & 140 & -50 & -30 & 5.94 \\
                 07:07 & 2356.25 & 3516.25 & 1305.00 & 32 & 150 & -60 & -30 & 11.98 \\
		\enddata
\end{deluxetable*}

\section{Summary and discussions} \label{sum}
In this paper, we first propose a geometrical model, featuring a torus-like flux rope based on the shape of 3DCORE model.
The global shape of the torus is an ellipse, while the cross section is circular along the torus.
The thinnest point is located between the Sun center and solar surface. Deflections and inclination of the torus are taken into account.
Using multiwavelength and multipoint observations from perspectives of Earth, STA, and SolO,
we apply the model to 3D reconstructions and tracking of the filament eruption for 35 minutes on 2024 October 8.
The filament, originating from AR 13847, was associated with a flare and a CME.
The morphology, direction, and true velocity ($\sim$433 km\,s$^{-1}$) of the eruptive filament are derived.
The filament propagates nonradially, deflecting slightly eastward by $\sim$10$\degr$ and significantly southward by $\sim$40$\degr$.
Trajectory of the filament in the ecliptic plane reveals that the filament heads for STA.
the true direction of the eruptive filament is the same using two independent methods, i.e., 3D reconstruction and spectral observation,
which confirms the reliability of our geometrical model.
The heliocentric distance of the filament increases from $\sim$1.68 to $\sim$2.94\,$R_{\sun}$ within 35 minutes.
According to the 3D reconstruction, the true speed of the CME leading edge is estimated to be 1046$-$1145 km\,s$^{-1}$, 
which is 2.5$\pm$0.1 times higher than that of filament. To our knowledge, this is the first report on the filament eruption on 2024 October 8.

The kinematic evolutions of CMEs and related filaments have been investigated substantially.
\citet{lsy25a} studied a sympathetic filament eruption triggered by an incoming EUV wave on 2024 February 9.
The filament eruption generates a slow CME propagating southward.
At 18:30 UT, the speeds of filament and associated CME are $\sim$68.5 and $\sim$155.5 km\,s$^{-1}$, respectively.
The ratio between the CME speed and filament speed reaches $\sim$2.27.
\citet{zqm25} applied the GCS model to 3D reconstruction of a long filament originating from farside of the Sun on 2023 March 12.
At 04:25 UT, the true speeds of filament and associated CME are $\sim$269 and $\sim$610 km\,s$^{-1}$, resulting in a ratio of $\sim$2.27.
\citet{zqm26} analyzed a fast CME driven by a nonradial filament eruption on 2023 December 31.
At 21:50 UT, the speeds of filament and CME front are $\sim$1476.4 and $\sim$2839 km\,s$^{-1}$, so that the ratio is close to $\sim$1.92.
\citet{lyn04} performed a 2.5D numerical simulation of a CME in the breakout model.
The synthetic coronagraphic images of CME present a typical three-part structure (see their Fig. 3).
The final velocities of CME front and bright core (filament) are 466.9 and 266.7 km\,s$^{-1}$, giving rise to a ratio of $\sim$1.75.
Similarly, in the 3D numerical simulation of a flux rope eruption and the related CME, the ratio between the speeds of CME front and core is $\sim$2.33 \citep{jia26}.
Both simulations reveal that CME fronts expand quickly during propagation, while the CME cores (filaments) expand negligibly.
Therefore, the speed of CME front is around twice higher than the speed of CME core (filament), which could reasonably explain our results.

It is emphasized that our geometrical model has limitations like other models. 
Firstly, the torus (flux rope) is coplanar and symmetric. Deformations as a result of writhing \citep{tor05} or skewing motions are not considered.
Secondly, magnetic field is not included as in the 3DCORE and FRi3D models. 
Prediction of the magnetic field strength and direction when the flux rope arrives at Earth or other planets is unavailable.
Thirdly, the adjustments of the parameters are manually rather than automatically. The best-fit model is determined subjectively.
Despite the above restrictions, our model could still be applied to 3D reconstruction of the filament on 2024 October 8.
The advantage of large FOVs of SCIUV and EUI allow for tracking the filament as far as $\sim$2.94\,$R_{\sun}$ at 05:55 UT,
when it is about to enter into the FOVs of WL coronagraphs (LASCO-C2 and STA/COR2).
Tracing the geometrical and kinematic evolutions of eruptive filaments uninterruptedly is of great importance to space weather forecast \citep{mie22,lsy25a}.
In the future, this geometrical model will be applied to more case studies and compared with previous methods, 
such as the GCS model, FRi3D model, and tie-pointing method. Comparisons are of great importance to validate and even improve our model.

So far, stereoscopic detections and tracking of solar eruptions have been listed as the primary scientific goals of in-orbit and forthcoming spacecrafts,
such as the multiview observatory for solar terrestrial science (MOST) \citep{gop24}, 
the Polarimeter to Unify the Corona and Heliosphere \citep[PUNCH;][]{def26}, ESA Vigil \citep{ame25}, 
the Lagrange-V Solar Observatory (LAVSO), and the Solar Polar orbit Observatory \citep[SPO;][]{deng25}. 
These missions will be located at L5 point (MOST, Vigil, LAVSO) or polar orbits (SPO). 
Using multipoint observations and a bundle of models, the 3D reconstructions will greatly improve our ability to predict space weather  
at Earth and other planets in the solar system \citep{pla16}.

\begin{acknowledgments}
The authors appreciate the reviewer for valuable suggestions to improve the quality of this article.
We also thank Dr. Xiaoyan Xie in Harvard-Smithsonian Center for Astrophysics for kind help and discussions.
SDO is a mission of NASA\rq{}s Living With a Star Program. HMI data are courtesy of the NASA/SDO science teams.
STEREO/SECCHI data are provided by a consortium of US, UK, Germany, Belgium, and France.
This CME catalog is generated and maintained at the CDAW Data Center by NASA and 
The Catholic University of America in cooperation with the Naval Research Laboratory. 
SOHO is a project of international cooperation between ESA and NASA.
Solar Orbiter is a space mission of international collaboration between ESA and NASA, operated by ESA. 
The Extreme Ultraviolet Imager (EUI) is part of the remote sensing instrument package of the ESA/NASA Solar Orbiter mission. 
SUVI was designed and built at Lockheed-Martin\rq{}s Advanced Technology Center in Palo Alto, California.
The CHASE mission is supported by the China National Space Administration (CNSA).
The ASO-S is supported by the Strategic Priority Research Program on Space Science, Chinese Academy of Sciences.
This work is supported by the Strategic Priority Research Program of the Chinese Academy of Sciences, grant No. XDB0560000,
the National Key R\&D Program of China 2021YFA1600500 (2021YFA1600502), 2022YFF0503002 (2022YFF0503000),
NSFC under the grant numbers 12373065, 12333009, 11790302, 12403064, 
Natural Science Foundation of Jiangsu Province (BK20231510, BK20241706),
and China\rq{}s Space Origins Exploration Program.
\end{acknowledgments}

\bibliography{recore3ds}

@ARTICLE{lem12,
	author = {{Lemen}, James R. and {Title}, Alan M. and {Akin}, David J. and {Boerner}, Paul F. and {Chou}, Catherine and {Drake}, Jerry F. and {Duncan}, Dexter W. and {Edwards}, Christopher G. and {Friedlaender}, Frank M. and {Heyman}, Gary F. and {Hurlburt}, Neal E. and {Katz}, Noah L. and {Kushner}, Gary D. and {Levay}, Michael and {Lindgren}, Russell W. and {Mathur}, Dnyanesh P. and {McFeaters}, Edward L. and {Mitchell}, Sarah and {Rehse}, Roger A. and {Schrijver}, Carolus J. and {Springer}, Larry A. and {Stern}, Robert A. and {Tarbell}, Theodore D. and {Wuelser}, Jean-Pierre and {Wolfson}, C. Jacob and {Yanari}, Carl and {Bookbinder}, Jay A. and {Cheimets}, Peter N. and {Caldwell}, David and {Deluca}, Edward E. and {Gates}, Richard and {Golub}, Leon and {Park}, Sang and {Podgorski}, William A. and {Bush}, Rock I. and {Scherrer}, Philip H. and {Gummin}, Mark A. and {Smith}, Peter and {Auker}, Gary and {Jerram}, Paul and {Pool}, Peter and {Soufli}, Regina and {Windt}, David L. and {Beardsley}, Sarah and {Clapp}, Matthew and {Lang}, James and {Waltham}, Nicholas},
	title = "{The Atmospheric Imaging Assembly (AIA) on the Solar Dynamics Observatory (SDO)}",
	journal = {\solphys},
	year = 2012,
	month = jan,
	volume = {275},
	number = {1-2},
	pages = {17-40},
	doi = {10.1007/s11207-011-9776-8},
	adsurl = {https://ui.adsabs.harvard.edu/abs/2012SoPh..275...17L}
}

@ARTICLE{sch12,
       author = {{Schou}, J. and {Scherrer}, P.~H. and {Bush}, R.~I. and {Wachter}, R. and {Couvidat}, S. and {Rabello-Soares}, M.~C. and {Bogart}, R.~S. and {Hoeksema}, J.~T. and {Liu}, Y. and {Duvall}, T.~L. and {Akin}, D.~J. and {Allard}, B.~A. and {Miles}, J.~W. and {Rairden}, R. and {Shine}, R.~A. and {Tarbell}, T.~D. and {Title}, A.~M. and {Wolfson}, C.~J. and {Elmore}, D.~F. and {Norton}, A.~A. and {Tomczyk}, S.},
        title = "{Design and Ground Calibration of the Helioseismic and Magnetic Imager (HMI) Instrument on the Solar Dynamics Observatory (SDO)}",
      journal = {\solphys},
         year = 2012,
        month = jan,
       volume = {275},
       number = {1-2},
        pages = {229-259},
          doi = {10.1007/s11207-011-9842-2},
       adsurl = {https://ui.adsabs.harvard.edu/abs/2012SoPh..275..229S}
}

@ARTICLE{pes12,
	author = {{Pesnell}, W. Dean and {Thompson}, B.~J. and {Chamberlin}, P.~C.},
	title = "{The Solar Dynamics Observatory (SDO)}",
	journal = {\solphys},
	year = 2012,
	month = jan,
	volume = {275},
	number = {1-2},
	pages = {3-15},
	doi = {10.1007/s11207-011-9841-3},
	adsurl = {https://ui.adsabs.harvard.edu/abs/2012SoPh..275....3P}
}

@ARTICLE{bru95,
	author = {{Brueckner}, G.~E. and {Howard}, R.~A. and {Koomen}, M.~J. and {Korendyke}, C.~M. and {Michels}, D.~J. and {Moses}, J.~D. and {Socker}, D.~G. and {Dere}, K.~P. and {Lamy}, P.~L. and {Llebaria}, A. and {Bout}, M.~V. and {Schwenn}, R. and {Simnett}, G.~M. and {Bedford}, D.~K. and {Eyles}, C.~J.},
	title = "{The Large Angle Spectroscopic Coronagraph (LASCO)}",
	journal = {\solphys},
	year = 1995,
	month = dec,
	volume = {162},
	number = {1-2},
	pages = {357-402},
	doi = {10.1007/BF00733434},
	adsurl = {https://ui.adsabs.harvard.edu/abs/1995SoPh..162..357B}
}

@ARTICLE{kai08,
	author = {{Kaiser}, M.~L. and {Kucera}, T.~A. and {Davila}, J.~M. and {St. Cyr}, O.~C. and {Guhathakurta}, M. and {Christian}, E.},
	title = "{The STEREO Mission: An Introduction}",
	journal = {\ssr},
	year = 2008,
	month = apr,
	volume = {136},
	number = {1-4},
	pages = {5-16},
	doi = {10.1007/s11214-007-9277-0},
	adsurl = {https://ui.adsabs.harvard.edu/abs/2008SSRv..136....5K}
}

@ARTICLE{how08,
	author = {{Howard}, R.~A. and {Moses}, J.~D. and {Vourlidas}, A. and {Newmark}, J.~S. and {Socker}, D.~G. and {Plunkett}, S.~P. and {Korendyke}, C.~M. and {Cook}, J.~W. and {Hurley}, A. and {Davila}, J.~M. and {Thompson}, W.~T. and {St Cyr}, O.~C. and {Mentzell}, E. and {Mehalick}, K. and {Lemen}, J.~R. and {Wuelser}, J.~P. and {Duncan}, D.~W. and {Tarbell}, T.~D. and {Wolfson}, C.~J. and {Moore}, A. and {Harrison}, R.~A. and {Waltham}, N.~R. and {Lang}, J. and {Davis}, C.~J. and {Eyles}, C.~J. and {Mapson-Menard}, H. and {Simnett}, G.~M. and {Halain}, J.~P. and {Defise}, J.~M. and {Mazy}, E. and {Rochus}, P. and {Mercier}, R. and {Ravet}, M.~F. and {Delmotte}, F. and {Auchere}, F. and {Delaboudiniere}, J.~P. and {Bothmer}, V. and {Deutsch}, W. and {Wang}, D. and {Rich}, N. and {Cooper}, S. and {Stephens}, V. and {Maahs}, G. and {Baugh}, R. and {McMullin}, D. and {Carter}, T.},
	title = "{Sun Earth Connection Coronal and Heliospheric Investigation (SECCHI)}",
	journal = {\ssr},
	year = 2008,
	month = apr,
	volume = {136},
	number = {1-4},
	pages = {67-115},
	doi = {10.1007/s11214-008-9341-4},
	adsurl = {https://ui.adsabs.harvard.edu/abs/2008SSRv..136...67H}
}

@ARTICLE{mu20,
       author = {{M{\"u}ller}, D. and {St. Cyr}, O.~C. and {Zouganelis}, I. and {Gilbert}, H.~R. and {Marsden}, R. and {Nieves-Chinchilla}, T. and {Antonucci}, E. and {Auch{\`e}re}, F. and {Berghmans}, D. and {Horbury}, T.~S. and {Howard}, R.~A. and {Krucker}, S. and {Maksimovic}, M. and {Owen}, C.~J. and {Rochus}, P. and {Rodriguez-Pacheco}, J. and {Romoli}, M. and {Solanki}, S.~K. and {Bruno}, R. and {Carlsson}, M. and {Fludra}, A. and {Harra}, L. and {Hassler}, D.~M. and {Livi}, S. and {Louarn}, P. and {Peter}, H. and {Sch{\"u}hle}, U. and {Teriaca}, L. and {del Toro Iniesta}, J.~C. and {Wimmer-Schweingruber}, R.~F. and {Marsch}, E. and {Velli}, M. and {De Groof}, A. and {Walsh}, A. and {Williams}, D.},
        title = "{The Solar Orbiter mission. Science overview}",
      journal = {\aap},
         year = 2020,
        month = oct,
       volume = {642},
          eid = {A1},
        pages = {A1},
          doi = {10.1051/0004-6361/202038467},
archivePrefix = {arXiv},
       eprint = {2009.00861},
 primaryClass = {astro-ph.SR},
       adsurl = {https://ui.adsabs.harvard.edu/abs/2020A&A...642A...1M}
}

@ARTICLE{ro20,
       author = {{Rochus}, P. and {Auch{\`e}re}, F. and {Berghmans}, D. and {Harra}, L. and {Schmutz}, W. and {Sch{\"u}hle}, U. and {Addison}, P. and {Appourchaux}, T. and {Aznar Cuadrado}, R. and {Baker}, D. and {Barbay}, J. and {Bates}, D. and {BenMoussa}, A. and {Bergmann}, M. and {Beurthe}, C. and {Borgo}, B. and {Bonte}, K. and {Bouzit}, M. and {Bradley}, L. and {B{\"u}chel}, V. and {Buchlin}, E. and {B{\"u}chner}, J. and {Cab{\'e}}, F. and {Cadiergues}, L. and {Chaigneau}, M. and {Chares}, B. and {Choque Cortez}, C. and {Coker}, P. and {Condamin}, M. and {Coumar}, S. and {Curdt}, W. and {Cutler}, J. and {Davies}, D. and {Davison}, G. and {Defise}, J.-M. and {Del Zanna}, G. and {Delmotte}, F. and {Delouille}, V. and {Dolla}, L. and {Dumesnil}, C. and {D{\"u}rig}, F. and {Enge}, R. and {Fran{\c{c}}ois}, S. and {Fourmond}, J.-J. and {Gillis}, J.-M. and {Giordanengo}, B. and {Gissot}, S. and {Green}, L.~M. and {Guerreiro}, N. and {Guilbaud}, A. and {Gyo}, M. and {Haberreiter}, M. and {Hafiz}, A. and {Hailey}, M. and {Halain}, J.-P. and {Hansotte}, J. and {Hecquet}, C. and {Heerlein}, K. and {Hellin}, M.-L. and {Hemsley}, S. and {Hermans}, A. and {Hervier}, V. and {Hochedez}, J.-F. and {Houbrechts}, Y. and {Ihsan}, K. and {Jacques}, L. and {J{\'e}r{\^o}me}, A. and {Jones}, J. and {Kahle}, M. and {Kennedy}, T. and {Klaproth}, M. and {Kolleck}, M. and {Koller}, S. and {Kotsialos}, E. and {Kraaikamp}, E. and {Langer}, P. and {Lawrenson}, A. and {Le Clech'}, J.-C. and {Lenaerts}, C. and {Liebecq}, S. and {Linder}, D. and {Long}, D.~M. and {Mampaey}, B. and {Markiewicz-Innes}, D. and {Marquet}, B. and {Marsch}, E. and {Matthews}, S. and {Mazy}, E. and {Mazzoli}, A. and {Meining}, S. and {Meltchakov}, E. and {Mercier}, R. and {Meyer}, S. and {Monecke}, M. and {Monfort}, F. and {Morinaud}, G. and {Moron}, F. and {Mountney}, L. and {M{\"u}ller}, R. and {Nicula}, B. and {Parenti}, S. and {Peter}, H. and {Pfiffner}, D. and {Philippon}, A. and {Phillips}, I. and {Plesseria}, J.-Y. and {Pylyser}, E. and {Rabecki}, F. and {Ravet-Krill}, M.-F. and {Rebellato}, J. and {Renotte}, E. and {Rodriguez}, L. and {Roose}, S. and {Rosin}, J. and {Rossi}, L. and {Roth}, P. and {Rouesnel}, F. and {Roulliay}, M. and {Rousseau}, A. and {Ruane}, K. and {Scanlan}, J. and {Schlatter}, P. and {Seaton}, D.~B. and {Silliman}, K. and {Smit}, S. and {Smith}, P.~J. and {Solanki}, S.~K. and {Spescha}, M. and {Spencer}, A. and {Stegen}, K. and {Stockman}, Y. and {Szwec}, N. and {Tamiatto}, C. and {Tandy}, J. and {Teriaca}, L. and {Theobald}, C. and {Tychon}, I. and {van Driel-Gesztelyi}, L. and {Verbeeck}, C. and {Vial}, J.-C. and {Werner}, S. and {West}, M.~J. and {Westwood}, D. and {Wiegelmann}, T. and {Willis}, G. and {Winter}, B. and {Zerr}, A. and {Zhang}, X. and {Zhukov}, A.~N.},
        title = "{The Solar Orbiter EUI instrument: The Extreme Ultraviolet Imager}",
      journal = {\aap},
         year = 2020,
        month = oct,
       volume = {642},
          eid = {A8},
        pages = {A8},
          doi = {10.1051/0004-6361/201936663},
       adsurl = {https://ui.adsabs.harvard.edu/abs/2020A&A...642A...8R}
}

@ARTICLE{kru20,
       author = {{Krucker}, S{\"a}m and {Hurford}, G.~J. and {Grimm}, O. and {K{\"o}gl}, S. and {Gr{\"o}belbauer}, H. -P. and {Etesi}, L. and {Casadei}, D. and {Csillaghy}, A. and {Benz}, A.~O. and {Arnold}, N.~G. and {Molendini}, F. and {Orleanski}, P. and {Schori}, D. and {Xiao}, H. and {Kuhar}, M. and {Hochmuth}, N. and {Felix}, S. and {Schramka}, F. and {Marcin}, S. and {Kobler}, S. and {Iseli}, L. and {Dreier}, M. and {Wiehl}, H.~J. and {Kleint}, L. and {Battaglia}, M. and {Lastufka}, E. and {Sathiapal}, H. and {Lapadula}, K. and {Bednarzik}, M. and {Birrer}, G. and {Stutz}, St. and {Wild}, Ch. and {Marone}, F. and {Skup}, K.~R. and {Cichocki}, A. and {Ber}, K. and {Rutkowski}, K. and {Bujwan}, W. and {Juchnikowski}, G. and {Winkler}, M. and {Darmetko}, M. and {Michalska}, M. and {Seweryn}, K. and {Bia{\l}ek}, A. and {Osica}, P. and {Sylwester}, J. and {Kowalinski}, M. and {{\'S}cis{\l}owski}, D. and {Siarkowski}, M. and {St{\k{e}}{\'s}licki}, M. and {Mrozek}, T. and {Podg{\'o}rski}, P. and {Meuris}, A. and {Limousin}, O. and {Gevin}, O. and {Le Mer}, I. and {Brun}, S. and {Strugarek}, A. and {Vilmer}, N. and {Musset}, S. and {Maksimovi{\'c}}, M. and {F{\'a}rn{\'\i}k}, F. and {Koz{\'a}{\v{c}}ek}, Z. and {Ka{\v{s}}parov{\'a}}, J. and {Mann}, G. and {{\"O}nel}, H. and {Warmuth}, A. and {Rendtel}, J. and {Anderson}, J. and {Bauer}, S. and {Dionies}, F. and {Paschke}, J. and {Pl{\"u}schke}, D. and {Woche}, M. and {Schuller}, F. and {Veronig}, A.~M. and {Dickson}, E.~C.~M. and {Gallagher}, P.~T. and {Maloney}, S.~A. and {Bloomfield}, D.~S. and {Piana}, M. and {Massone}, A.~M. and {Benvenuto}, F. and {Massa}, P. and {Schwartz}, R.~A. and {Dennis}, B.~R. and {van Beek}, H.~F. and {Rodr{\'\i}guez-Pacheco}, J. and {Lin}, R.~P.},
        title = "{The Spectrometer/Telescope for Imaging X-rays (STIX)}",
      journal = {\aap},
         year = 2020,
        month = oct,
       volume = {642},
          eid = {A15},
        pages = {A15},
          doi = {10.1051/0004-6361/201937362},
       adsurl = {https://ui.adsabs.harvard.edu/abs/2020A&A...642A..15K}
}

@ARTICLE{chd25,
       author = {{Chen}, Huadong and {Xia}, Chun and {Ma}, Suli and {Su}, Yingna and {Zhou}, Guiping and {Priest}, Eric and {Fletcher}, Lyndsay and {Shen}, Yuandeng and {Tu}, Weining and {Wang}, Wei and {Zhang}, Jun},
        title = "{Magnetic Dip Found in a Quiescent Prominence Foot via Observation and Simulation}",
      journal = {\apj},
         year = 2025,
        month = nov,
       volume = {994},
       number = {1},
          eid = {27},
        pages = {27},
          doi = {10.3847/1538-4357/ae0ad4},
archivePrefix = {arXiv},
       eprint = {2509.19723},
 primaryClass = {astro-ph.SR},
       adsurl = {https://ui.adsabs.harvard.edu/abs/2025ApJ...994...27C}
}

@ARTICLE{chd24,
       author = {{Chen}, Huadong and {Fletcher}, Lyndsay and {Zhou}, Guiping and {Cheng}, Xin and {Wang}, Ya and {Mulay}, Sargam and {Zheng}, Ruisheng and {Ma}, Suli and {Zhang}, Xiaofan},
        title = "{Simultaneous Eruption and Shrinkage of Preexisting Flare Loops during a Subsequent Solar Eruption}",
      journal = {\apj},
         year = 2024,
        month = dec,
       volume = {976},
       number = {2},
          eid = {207},
        pages = {207},
          doi = {10.3847/1538-4357/ad8c25},
archivePrefix = {arXiv},
       eprint = {2410.12202},
 primaryClass = {astro-ph.SR},
       adsurl = {https://ui.adsabs.harvard.edu/abs/2024ApJ...976..207C}
}

@ARTICLE{chc25,
       author = {{Chen}, Hechao and {Tian}, Hui and {Zhang}, Quanhao and {Li}, Chuan and {Xia}, Chun and {Bai}, Xianyong and {Hou}, Zhenyong and {Ji}, Kaifan and {Deng}, Yuanyong and {Yang}, Xiao and {Hu}, Ziyao},
        title = "{Minifilament Eruptions as the Last Straw to Break the Equilibrium of a Giant Solar Filament}",
      journal = {\apj},
         year = 2025,
        month = apr,
       volume = {983},
       number = {2},
          eid = {143},
        pages = {143},
          doi = {10.3847/1538-4357/adc12a},
archivePrefix = {arXiv},
       eprint = {2503.13800},
 primaryClass = {astro-ph.SR},
       adsurl = {https://ui.adsabs.harvard.edu/abs/2025ApJ...983..143C}
}

@ARTICLE{gyh25,
       author = {{Gao}, Yuhang and {Tian}, Hui and {Berghmans}, David and {Duan}, Yadan and {Van Doorsselaere}, Tom and {Chen}, Hechao and {Kraaikamp}, Emil},
        title = "{Reconnection Nanojets in an Erupting Solar Filament with Unprecedented High Speeds}",
      journal = {\apjl},
         year = 2025,
        month = may,
       volume = {985},
       number = {1},
          eid = {L12},
        pages = {L12},
          doi = {10.3847/2041-8213/add33a},
archivePrefix = {arXiv},
       eprint = {2504.20663},
 primaryClass = {astro-ph.SR},
       adsurl = {https://ui.adsabs.harvard.edu/abs/2025ApJ...985L..12G}
}

@ARTICLE{bi13,
       author = {{Bi}, Yi and {Jiang}, Yunchun and {Yang}, Jiayan and {Zheng}, Ruisheng and {Hong}, Junchao and {Li}, Haidong and {Yang}, Dan and {Yang}, Bo},
        title = "{Analysis of the Simultaneous Rotation and Non-radial Propagation of an Eruptive Filament}",
      journal = {\apj},
         year = 2013,
        month = aug,
       volume = {773},
       number = {2},
          eid = {162},
        pages = {162},
          doi = {10.1088/0004-637X/773/2/162},
       adsurl = {https://ui.adsabs.harvard.edu/abs/2013ApJ...773..162B}
}

@ARTICLE{ant21,
       author = {{Antolin}, Patrick and {Pagano}, Paolo and {Testa}, Paola and {Petralia}, Antonino and {Reale}, Fabio},
        title = "{Reconnection nanojets in the solar corona}",
      journal = {Nature Astronomy},
         year = 2021,
        month = jan,
       volume = {5},
        pages = {54-62},
          doi = {10.1038/s41550-020-1199-8},
       adsurl = {https://ui.adsabs.harvard.edu/abs/2021NatAs...5...54A}
}

@ARTICLE{tor05,
       author = {{T{\"o}r{\"o}k}, T. and {Kliem}, B.},
        title = "{Confined and Ejective Eruptions of Kink-unstable Flux Ropes}",
      journal = {\apjl},
         year = 2005,
        month = sep,
       volume = {630},
       number = {1},
        pages = {L97-L100},
          doi = {10.1086/462412},
archivePrefix = {arXiv},
       eprint = {astro-ph/0507662},
 primaryClass = {astro-ph},
       adsurl = {https://ui.adsabs.harvard.edu/abs/2005ApJ...630L..97T}
}

@ARTICLE{shq22,
       author = {{Song}, Hongqiang and {Li}, Leping and {Chen}, Yao},
        title = "{Toward a Unified Explanation for the Three-part Structure of Solar Coronal Mass Ejections}",
      journal = {\apj},
         year = 2022,
        month = jul,
       volume = {933},
       number = {1},
          eid = {68},
        pages = {68},
          doi = {10.3847/1538-4357/ac7239},
archivePrefix = {arXiv},
       eprint = {2205.11682},
 primaryClass = {astro-ph.SR},
       adsurl = {https://ui.adsabs.harvard.edu/abs/2022ApJ...933...68S}
}

@ARTICLE{cx20,
       author = {{Cheng}, X. and {Zhang}, J. and {Kliem}, B. and {T{\"o}r{\"o}k}, T. and {Xing}, C. and {Zhou}, Z.~J. and {Inhester}, B. and {Ding}, M.~D.},
        title = "{Initiation and Early Kinematic Evolution of Solar Eruptions}",
      journal = {\apj},
         year = 2020,
        month = may,
       volume = {894},
       number = {2},
          eid = {85},
        pages = {85},
          doi = {10.3847/1538-4357/ab886a},
archivePrefix = {arXiv},
       eprint = {2004.03790},
 primaryClass = {astro-ph.SR},
       adsurl = {https://ui.adsabs.harvard.edu/abs/2020ApJ...894...85C}
}

@ARTICLE{xing25,
       author = {{Xing}, Chen and {Cheng}, Xin and {Aulanier}, Guillaume and {Ding}, Mingde},
        title = "{Initiation Route of Coronal Mass Ejections. II. The Role of Filament Mass}",
      journal = {\apj},
         year = 2025,
        month = jun,
       volume = {986},
       number = {1},
          eid = {37},
        pages = {37},
          doi = {10.3847/1538-4357/adceb5},
archivePrefix = {arXiv},
       eprint = {2504.14876},
 primaryClass = {astro-ph.SR},
       adsurl = {https://ui.adsabs.harvard.edu/abs/2025ApJ...986...37X}
}

@ARTICLE{zqm25,
       author = {{Zhang}, Qingmin and {Pan}, Wenwei and {Ying}, Beili and {Feng}, Li and {Li}, Yiliang and {Yan}, Xiaoli and {Yang}, Liheng and {Qiu}, Ye and {Chen}, Jun and {Ma}, Suli},
        title = "{Tracking an Eruptive Intermediate Prominence Originating from the Farside of the Sun}",
      journal = {\apj},
         year = 2025,
        month = jun,
       volume = {985},
       number = {2},
          eid = {237},
        pages = {237},
          doi = {10.3847/1538-4357/add328},
archivePrefix = {arXiv},
       eprint = {2505.01684},
 primaryClass = {astro-ph.SR},
       adsurl = {https://ui.adsabs.harvard.edu/abs/2025ApJ...985..237Z}
}

@ARTICLE{zqm26,
       author = {{Zhang}, Qingmin and {Ning}, Zongjun and {Chen}, Xingyao and {Chen}, Wei and {Yan}, Xiaoli and {Li}, Shuyue},
        title = "{Herringbone Structures during an X-class Eruptive Flare}",
      journal = {\apj},
         year = 2026,
        month = mar,
       volume = {999},
       number = {1},
          eid = {66},
        pages = {66},
          doi = {10.3847/1538-4357/ae4024},
archivePrefix = {arXiv},
       eprint = {2602.01117},
 primaryClass = {astro-ph.SR},
       adsurl = {https://ui.adsabs.harvard.edu/abs/2026ApJ...999...66Z}
}

@ARTICLE{zqm24,
       author = {{Zhang}, Qingmin and {Ou}, Yudi and {Huang}, Zhenghua and {Song}, Yongliang and {Ma}, Suli},
        title = "{Tracking an Eruptive Prominence Using Multiwavelength and Multiview Observations on 2023 March 7}",
      journal = {\apj},
         year = 2024,
        month = dec,
       volume = {977},
       number = {1},
          eid = {4},
        pages = {4},
          doi = {10.3847/1538-4357/ad8bad},
archivePrefix = {arXiv},
       eprint = {2410.22724},
 primaryClass = {astro-ph.SR},
       adsurl = {https://ui.adsabs.harvard.edu/abs/2024ApJ...977....4Z}
}

@ARTICLE{zqm12,
       author = {{Zhang}, Q.~M. and {Chen}, P.~F. and {Xia}, C. and {Keppens}, R.},
        title = "{Observations and simulations of longitudinal oscillations of an active region prominence}",
      journal = {\aap},
         year = 2012,
        month = jun,
       volume = {542},
          eid = {A52},
        pages = {A52},
          doi = {10.1051/0004-6361/201218786},
archivePrefix = {arXiv},
       eprint = {1204.3787},
 primaryClass = {astro-ph.SR},
       adsurl = {https://ui.adsabs.harvard.edu/abs/2012A&A...542A..52Z}
}

@ARTICLE{zqm23,
       author = {{Zhang}, Qing-Min and {Hou}, Zhen-Yong and {Bai}, Xian-Yong},
        title = "{A Revised Graduated Cylindrical Shell Model and its Application to a Prominence Eruption}",
      journal = {Research in Astronomy and Astrophysics},
         year = 2023,
        month = dec,
       volume = {23},
       number = {12},
          eid = {125004},
        pages = {125004},
          doi = {10.1088/1674-4527/acee4d},
archivePrefix = {arXiv},
       eprint = {2307.00943},
 primaryClass = {astro-ph.SR},
       adsurl = {https://ui.adsabs.harvard.edu/abs/2023RAA....23l5004Z}
}

@ARTICLE{mie22,
       author = {{Mierla}, M. and {Zhukov}, A.~N. and {Berghmans}, D. and {Parenti}, S. and {Auch{\`e}re}, F. and {Heinzel}, P. and {Seaton}, D.~B. and {Palmerio}, E. and {Jej{\v{c}}i{\v{c}}}, S. and {Janssens}, J. and {Kraaikamp}, E. and {Nicula}, B. and {Long}, D.~M. and {Hayes}, L.~A. and {Jebaraj}, I.~C. and {Talpeanu}, D.-C. and {D'Huys}, E. and {Dolla}, L. and {Gissot}, S. and {Magdaleni{\'c}}, J. and {Rodriguez}, L. and {Shestov}, S. and {Stegen}, K. and {Verbeeck}, C. and {Sasso}, C. and {Romoli}, M. and {Andretta}, V.},
        title = "{Prominence eruption observed in He II 304 {\r{A}} up to >6 R$_{{\ensuremath{\odot}}}$ by EUI/FSI aboard Solar Orbiter}",
      journal = {\aap},
         year = 2022,
        month = jun,
       volume = {662},
          eid = {L5},
        pages = {L5},
          doi = {10.1051/0004-6361/202244020},
archivePrefix = {arXiv},
       eprint = {2205.15214},
 primaryClass = {astro-ph.SR},
       adsurl = {https://ui.adsabs.harvard.edu/abs/2022A&A...662L...5M}
}

@ARTICLE{liu12,
       author = {{Liu}, Wei and {Berger}, Thomas E. and {Low}, B.~C.},
        title = "{First SDO/AIA Observation of Solar Prominence Formation Following an Eruption: Magnetic Dips and Sustained Condensation and Drainage}",
      journal = {\apjl},
         year = 2012,
        month = feb,
       volume = {745},
       number = {2},
          eid = {L21},
        pages = {L21},
          doi = {10.1088/2041-8205/745/2/L21},
       adsurl = {https://ui.adsabs.harvard.edu/abs/2012ApJ...745L..21L}
}

@ARTICLE{aul98,
       author = {{Aulanier}, G. and {Demoulin}, P. and {van Driel-Gesztelyi}, L. and {Mein}, P. and {Deforest}, C.},
        title = "{3-D magnetic configurations supporting prominences. II. The lateral feet as a perturbation of a twisted flux-tube}",
      journal = {\aap},
         year = 1998,
        month = jul,
       volume = {335},
        pages = {309-322},
       adsurl = {https://ui.adsabs.harvard.edu/abs/1998A&A...335..309A}
}

@ARTICLE{ste16,
       author = {{Sterling}, Alphonse C. and {Moore}, Ronald L. and {Falconer}, David A. and {Panesar}, Navdeep K. and {Akiyama}, Sachiko and {Yashiro}, Seiji and {Gopalswamy}, Nat},
        title = "{Minifilament Eruptions that Drive Coronal Jets in a Solar Active Region}",
      journal = {\apj},
         year = 2016,
        month = apr,
       volume = {821},
       number = {2},
          eid = {100},
        pages = {100},
          doi = {10.3847/0004-637X/821/2/100},
       adsurl = {https://ui.adsabs.harvard.edu/abs/2016ApJ...821..100S}
}

@ARTICLE{mie10,
       author = {{Mierla}, M. and {Inhester}, B. and {Antunes}, A. and {Boursier}, Y. and {Byrne}, J.~P. and {Colaninno}, R. and {Davila}, J. and {de Koning}, C.~A. and {Gallagher}, P.~T. and {Gissot}, S. and {Howard}, R.~A. and {Howard}, T.~A. and {Kramar}, M. and {Lamy}, P. and {Liewer}, P.~C. and {Maloney}, S. and {Marqu{\'e}}, C. and {McAteer}, R.~T.~J. and {Moran}, T. and {Rodriguez}, L. and {Srivastava}, N. and {St. Cyr}, O.~C. and {Stenborg}, G. and {Temmer}, M. and {Thernisien}, A. and {Vourlidas}, A. and {West}, M.~J. and {Wood}, B.~E. and {Zhukov}, A.~N.},
        title = "{On the 3-D reconstruction of Coronal Mass Ejections using coronagraph data}",
      journal = {Annales Geophysicae},
         year = 2010,
        month = jan,
       volume = {28},
       number = {1},
        pages = {203-215},
          doi = {10.5194/angeo-28-203-2010},
       adsurl = {https://ui.adsabs.harvard.edu/abs/2010AnGeo..28..203M}
}

@ARTICLE{kay24,
       author = {{Kay}, C. and {Palmerio}, E.},
        title = "{Collection, Collation, and Comparison of 3D Coronal CME Reconstructions}",
      journal = {Space Weather},
         year = 2024,
        month = jan,
       volume = {22},
       number = {1},
          eid = {e2023SW003796},
        pages = {e2023SW003796},
          doi = {10.1029/2023SW003796},
archivePrefix = {arXiv},
       eprint = {2311.10712},
 primaryClass = {astro-ph.SR},
       adsurl = {https://ui.adsabs.harvard.edu/abs/2024SpWea..2203796K}
}

@ARTICLE{mu25,
       author = {{M{\"u}ller}, Daniel and {Ireland}, Jack and {De Groof}, Anik and {Dimitoglou}, George and {Fleck}, Bernhard},
        title = "{SOHO's 30-year legacy of observing the Sun}",
      journal = {Nature Astronomy},
         year = 2025,
        month = dec,
          doi = {10.1038/s41550-025-02687-4},
       adsurl = {https://ui.adsabs.harvard.edu/abs/2025NatAs.tmp..233M}
}

@ARTICLE{lia25,
       author = {{Liakh}, Valeriia and {Jenkins}, Jack},
        title = "{Numerical Modeling of Prominences and Coronal Rain with the MPI-AMRVAC Code}",
      journal = {\solphys},
         year = 2025,
        month = oct,
       volume = {300},
       number = {10},
          eid = {147},
        pages = {147},
          doi = {10.1007/s11207-025-02552-7},
archivePrefix = {arXiv},
       eprint = {2510.04222},
 primaryClass = {astro-ph.SR},
       adsurl = {https://ui.adsabs.harvard.edu/abs/2025SoPh..300..147L}
}

@ARTICLE{mo14,
       author = {{M{\"o}stl}, C. and {Amla}, K. and {Hall}, J.~R. and {Liewer}, P.~C. and {De Jong}, E.~M. and {Colaninno}, R.~C. and {Veronig}, A.~M. and {Rollett}, T. and {Temmer}, M. and {Peinhart}, V. and {Davies}, J.~A. and {Lugaz}, N. and {Liu}, Y.~D. and {Farrugia}, C.~J. and {Luhmann}, J.~G. and {Vr{\v{s}}nak}, B. and {Harrison}, R.~A. and {Galvin}, A.~B.},
        title = "{Connecting Speeds, Directions and Arrival Times of 22 Coronal Mass Ejections from the Sun to 1 AU}",
      journal = {\apj},
         year = 2014,
        month = jun,
       volume = {787},
       number = {2},
          eid = {119},
        pages = {119},
          doi = {10.1088/0004-637X/787/2/119},
archivePrefix = {arXiv},
       eprint = {1404.3579},
 primaryClass = {astro-ph.SR},
       adsurl = {https://ui.adsabs.harvard.edu/abs/2014ApJ...787..119M}
}

@ARTICLE{mo15,
       author = {{M{\"o}stl}, Christian and {Rollett}, Tanja and {Frahm}, Rudy A. and {Liu}, Ying D. and {Long}, David M. and {Colaninno}, Robin C. and {Reiss}, Martin A. and {Temmer}, Manuela and {Farrugia}, Charles J. and {Posner}, Arik and {Dumbovi{\'c}}, Mateja and {Janvier}, Miho and {D{\'e}moulin}, Pascal and {Boakes}, Peter and {Devos}, Andy and {Kraaikamp}, Emil and {Mays}, Mona L. and {Vr{\v{s}}nak}, Bojan},
        title = "{Strong coronal channelling and interplanetary evolution of a solar storm up to Earth and Mars}",
      journal = {Nature Communications},
         year = 2015,
        month = may,
       volume = {6},
          eid = {7135},
        pages = {7135},
          doi = {10.1038/ncomms8135},
archivePrefix = {arXiv},
       eprint = {1506.02842},
 primaryClass = {astro-ph.SR},
       adsurl = {https://ui.adsabs.harvard.edu/abs/2015NatCo...6.7135M}
}

@ARTICLE{mo18,
       author = {{M{\"o}stl}, C. and {Amerstorfer}, T. and {Palmerio}, E. and {Isavnin}, A. and {Farrugia}, C.~J. and {Lowder}, C. and {Winslow}, R.~M. and {Donnerer}, J.~M. and {Kilpua}, E.~K.~J. and {Boakes}, P.~D.},
        title = "{Forward Modeling of Coronal Mass Ejection Flux Ropes in the Inner Heliosphere with 3DCORE}",
      journal = {Space Weather},
         year = 2018,
        month = mar,
       volume = {16},
       number = {3},
        pages = {216-229},
          doi = {10.1002/2017SW001735},
archivePrefix = {arXiv},
       eprint = {1710.00587},
 primaryClass = {astro-ph.SR},
       adsurl = {https://ui.adsabs.harvard.edu/abs/2018SpWea..16..216M}
}

@ARTICLE{rud24,
       author = {{R{\"u}disser}, Hannah T. and {Weiss}, Andreas J. and {Le Lou{\"e}dec}, Justin and {Amerstorfer}, Ute V. and {M{\"o}stl}, Christian and {Davies}, Emma E. and {Lammer}, Helmut},
        title = "{Understanding the Effects of Spacecraft Trajectories through Solar Coronal Mass Ejection Flux Ropes Using 3DCOREweb}",
      journal = {\apj},
         year = 2024,
        month = oct,
       volume = {973},
       number = {2},
          eid = {150},
        pages = {150},
          doi = {10.3847/1538-4357/ad660a},
archivePrefix = {arXiv},
       eprint = {2405.03271},
 primaryClass = {astro-ph.SR},
       adsurl = {https://ui.adsabs.harvard.edu/abs/2024ApJ...973..150R}
}

@ARTICLE{wei21a,
       author = {{Weiss}, Andreas J. and {M{\"o}stl}, Christian and {Amerstorfer}, Tanja and {Bailey}, Rachel L. and {Reiss}, Martin A. and {Hinterreiter}, J{\"u}rgen and {Amerstorfer}, Ute A. and {Bauer}, Maike},
        title = "{Analysis of Coronal Mass Ejection Flux Rope Signatures Using 3DCORE and Approximate Bayesian Computation}",
      journal = {\apjs},
         year = 2021,
        month = jan,
       volume = {252},
       number = {1},
          eid = {9},
        pages = {9},
          doi = {10.3847/1538-4365/abc9bd},
archivePrefix = {arXiv},
       eprint = {2009.00327},
 primaryClass = {astro-ph.SR},
       adsurl = {https://ui.adsabs.harvard.edu/abs/2021ApJS..252....9W}
}

@ARTICLE{wei21b,
       author = {{Weiss}, A.~J. and {M{\"o}stl}, C. and {Davies}, E.~E. and {Amerstorfer}, T. and {Bauer}, M. and {Hinterreiter}, J. and {Reiss}, M.~A. and {Bailey}, R.~L. and {Horbury}, T.~S. and {O'Brien}, H. and {Evans}, V. and {Angelini}, V. and {Heyner}, D. and {Richter}, I. and {Auster}, H.-U. and {Magnes}, W. and {Fischer}, D. and {Baumjohann}, W.},
        title = "{Multi-point analysis of coronal mass ejection flux ropes using combined data from Solar Orbiter, BepiColombo, and Wind}",
      journal = {\aap},
         year = 2021,
        month = dec,
       volume = {656},
          eid = {A13},
        pages = {A13},
          doi = {10.1051/0004-6361/202140919},
archivePrefix = {arXiv},
       eprint = {2103.16187},
 primaryClass = {astro-ph.SR},
       adsurl = {https://ui.adsabs.harvard.edu/abs/2021A&A...656A..13W}
}

@ARTICLE{kw14,
       author = {{Kwon}, Ryun-Young and {Zhang}, Jie and {Olmedo}, Oscar},
        title = "{New Insights into the Physical Nature of Coronal Mass Ejections and Associated Shock Waves within the Framework of the Three-dimensional Structure}",
      journal = {\apj},
         year = 2014,
        month = oct,
       volume = {794},
       number = {2},
          eid = {148},
        pages = {148},
          doi = {10.1088/0004-637X/794/2/148},
       adsurl = {https://ui.adsabs.harvard.edu/abs/2014ApJ...794..148K}
}

@ARTICLE{ver23,
       author = {{Verbeke}, Christine and {Mays}, M. Leila and {Kay}, Christina and {Riley}, Pete and {Palmerio}, Erika and {Dumbovi{\'c}}, Mateja and {Mierla}, Marilena and {Scolini}, Camilla and {Temmer}, Manuela and {Paouris}, Evangelos and {Balmaceda}, Laura A. and {Cremades}, Hebe and {Hinterreiter}, J{\"u}rgen},
        title = "{Quantifying errors in 3D CME parameters derived from synthetic data using white-light reconstruction techniques}",
      journal = {Advances in Space Research},
         year = 2023,
        month = dec,
       volume = {72},
       number = {12},
        pages = {5243-5262},
          doi = {10.1016/j.asr.2022.08.056},
archivePrefix = {arXiv},
       eprint = {2302.00531},
 primaryClass = {astro-ph.SR},
       adsurl = {https://ui.adsabs.harvard.edu/abs/2023AdSpR..72.5243V}
}

@ARTICLE{hhd26,
       author = {{Hu}, Huidong and {Chen}, Chong and {Jiao}, Yiming and {Zhu}, Bei and {Wang}, Rui and {Zhao}, Xiaowei and {Yang}, Liping},
        title = "{Lateral Deformation of Large-scale Coronal Mass Ejections during the Transition from Nonradial to Radial Propagation}",
      journal = {\apj},
         year = 2026,
        month = feb,
       volume = {997},
       number = {2},
          eid = {303},
        pages = {303},
          doi = {10.3847/1538-4357/ae267e},
archivePrefix = {arXiv},
       eprint = {2512.09937},
 primaryClass = {physics.space-ph},
       adsurl = {https://ui.adsabs.harvard.edu/abs/2026ApJ...997..303H}
}

@ARTICLE{is16,
       author = {{Isavnin}, A.},
        title = "{FRiED: A Novel Three-dimensional Model of Coronal Mass Ejections}",
      journal = {\apj},
         year = 2016,
        month = dec,
       volume = {833},
       number = {2},
          eid = {267},
        pages = {267},
          doi = {10.3847/1538-4357/833/2/267},
archivePrefix = {arXiv},
       eprint = {1703.01659},
 primaryClass = {physics.space-ph},
       adsurl = {https://ui.adsabs.harvard.edu/abs/2016ApJ...833..267I}
}

@ARTICLE{chen22,
       author = {{Chen}, Yuhao and {Ye}, Jing and {Mei}, Zhixing and {Shen}, Chengcai and {Roussev}, Ilia I. and {Forbes}, Terry G. and {Lin}, Jun and {Ziegler}, Udo},
        title = "{Numerical Investigations of Catastrophe in Coronal Magnetic Configuration Triggered by Newly Emerging Flux}",
      journal = {\apj},
         year = 2022,
        month = jul,
       volume = {933},
       number = {2},
          eid = {148},
        pages = {148},
          doi = {10.3847/1538-4357/ac73ef},
       adsurl = {https://ui.adsabs.harvard.edu/abs/2022ApJ...933..148C}
}

@ARTICLE{zqm22,
       author = {{Zhang}, Q.~M.},
        title = "{Tracking the 3D evolution of a halo coronal mass ejection using the revised cone model}",
      journal = {\aap},
         year = 2022,
        month = apr,
       volume = {660},
          eid = {A144},
        pages = {A144},
          doi = {10.1051/0004-6361/202142942},
archivePrefix = {arXiv},
       eprint = {2202.10676},
 primaryClass = {astro-ph.SR},
       adsurl = {https://ui.adsabs.harvard.edu/abs/2022A&A...660A.144Z}
}

@ARTICLE{lxh25,
       author = {{Li}, Xiaohong and {Solanki}, Sami K. and {Wiegelmann}, Thomas and {Valori}, Gherardo and {Calchetti}, Daniele and {Hirzberger}, Johann and {Dur{\'a}n}, Juan Sebasti{\'a}n Castellanos and {Woch}, Joachim and {Gandorfer}, Achim and {the Solar Orbiter team}},
        title = "{3D structures of the base of small-scale recurrent jets revealed by Solar Orbiter}",
      journal = {\aap},
         year = 2025,
        month = oct,
       volume = {702},
          eid = {A201},
        pages = {A201},
          doi = {10.1051/0004-6361/202555972},
archivePrefix = {arXiv},
       eprint = {2509.06792},
 primaryClass = {astro-ph.SR},
       adsurl = {https://ui.adsabs.harvard.edu/abs/2025A&A...702A.201L}
}

@ARTICLE{dyd26,
       author = {{Duan}, Yadan and {Yan}, Xiaoli and {Hong}, Junchao and {Chen}, Hechao and {Gao}, Yuhang and {Sun}, Zheng and {Hou}, Zhenyong and {Wang}, Jincheng},
        title = "{Resolving interchange reconnection dynamics in a fan-spine-like topology observed by Solar Orbiter}",
      journal = {\aap},
         year = 2026,
        month = jan,
       volume = {706},
          eid = {A1},
        pages = {A1},
          doi = {10.1051/0004-6361/202557158},
archivePrefix = {arXiv},
       eprint = {2512.01209},
 primaryClass = {astro-ph.SR},
       adsurl = {https://ui.adsabs.harvard.edu/abs/2026A&A...706A...1D}
}

@ARTICLE{lep90,
       author = {{Lepping}, R.~P. and {Jones}, J.~A. and {Burlaga}, L.~F.},
        title = "{Magnetic field structure of interplanetary magnetic clouds at 1 AU}",
      journal = {\jgr},
         year = 1990,
        month = aug,
       volume = {95},
       number = {A8},
        pages = {11957-11965},
          doi = {10.1029/JA095iA08p11957},
       adsurl = {https://ui.adsabs.harvard.edu/abs/1990JGR....9511957L}
}

@ARTICLE{zur06,
       author = {{Zurbuchen}, Thomas H. and {Richardson}, Ian G.},
        title = "{In-Situ Solar Wind and Magnetic Field Signatures of Interplanetary Coronal Mass Ejections}",
      journal = {\ssr},
         year = 2006,
        month = mar,
       volume = {123},
       number = {1-3},
        pages = {31-43},
          doi = {10.1007/s11214-006-9010-4},
       adsurl = {https://ui.adsabs.harvard.edu/abs/2006SSRv..123...31Z}
}

@ARTICLE{jan15,
       author = {{Janvier}, M. and {Aulanier}, G. and {D{\'e}moulin}, P.},
        title = "{From Coronal Observations to MHD Simulations, the Building Blocks for 3D Models of Solar Flares (Invited Review)}",
      journal = {\solphys},
         year = 2015,
        month = dec,
       volume = {290},
       number = {12},
        pages = {3425-3456},
          doi = {10.1007/s11207-015-0710-3},
archivePrefix = {arXiv},
       eprint = {1505.05299},
 primaryClass = {astro-ph.SR},
       adsurl = {https://ui.adsabs.harvard.edu/abs/2015SoPh..290.3425J}
}

@ARTICLE{def26,
       author = {{DeForest}, Craig E. and {Gibson}, Sarah E. and {Killough}, Ronnie and {Waltham}, Nick R. and {Beasley}, Matt N. and {Colaninno}, Robin C. and {Laurent}, Glenn T. and {Seaton}, Daniel B. and {Hughes}, J. Marcus and {Guhathakurta}, Madhulika and {Viall}, Nicholeen M. and {Atti{\'e}}, Raphael and {Banerjee}, Dipankar and {Barnard}, Luke and {Biesecker}, Doug A. and {Bisi}, Mario M. and {Bothmer}, Volker and {Brody}, Antonina and {Burkepile}, Joan and {Cairns}, Iver H. and {Campbell}, Jennifer L. and {Case}, Traci R. and {Caspi}, Amir and {Cheney}, David and {Chhiber}, Rohit and {Clapp}, Matthew J. and {Cranmer}, Steven R. and {Davies}, Jackie A. and {de Koning}, Curt A. and {Desai}, Mihir I. and {Elliott}, Heather A. and {Farid}, Samaiyah and {Gallardo-Lacourt}, Bea and {Gilly}, Chris and {Gobat}, Caden and {Hanson}, Mary H. and {Harrison}, Richard A. and {Hassler}, Donald M. and {Henley}, Chase and {Henry}, Alan M. and {Howard}, Russell A. and {Jackson}, Bernard V. and {Jones}, Samuel and {Kolinski}, Don and {Lamb}, Derek A. and {Lehtinen}, Florine and {Lowder}, Chris and {Malanushenko}, Anna and {Matthaeus}, William H. and {McComas}, David J. and {McGee}, Jacob and {Morgan}, Huw and {Oberoi}, Divya and {Odstrcil}, Dusan and {Parmenter}, Chris and {Patel}, Ritesh and {Pecora}, Francesco and {Persyn}, Steve and {Pizzo}, Victor J. and {Plunkett}, Simon P. and {Provornikova}, Elena and {Raouafi}, Nour Eddine and {Redfern}, Jillian A. and {Rouillard}, Alexis P. and {Smith}, Kelly D. and {Smith}, Keith B. and {Talpas}, Zachary S. and {Tappin}, S. James and {Thernisien}, Arnaud and {Thompson}, Barbara J. and {Van Kooten}, Samuel and {Walsh}, Kevin J. and {Webb}, David F. and {Wells}, William L. and {West}, Matthew J. and {Wiens}, Zachary and {Yang}, Yan and {Zhukov}, Andrei N.},
        title = "{Polarimeter to Unify the Corona and Heliosphere (PUNCH)}",
      journal = {\solphys},
         year = 2026,
        month = jan,
       volume = {301},
       number = {1},
          eid = {16},
        pages = {16},
          doi = {10.1007/s11207-026-02608-2},
archivePrefix = {arXiv},
       eprint = {2509.15131},
 primaryClass = {astro-ph.SR},
       adsurl = {https://ui.adsabs.harvard.edu/abs/2026SoPh..301...16D}
}

@ARTICLE{deng25,
       author = {{Deng}, Yuanyong and {Tian}, Hui and {Jiang}, Jie and {Yang}, Shuhong and {Li}, Hao and {Cameron}, Robert and {Gizon}, Laurent and {Harra}, Louise and {Wimmer-Schweingruber}, Robert F. and {Auch{\`e}re}, Fr{\'e}d{\'e}ric and {Bai}, Xianyong and {Bellot}, Rubio Luis and {Chen}, Linjie and {Chen}, Pengfei and {Chitta}, Lakshmi Pradeep and {Davies}, Jackie and {Favata}, Fabio and {Feng}, Li and {Feng}, Xueshang and {Gan}, Weiqun and {Hassler}, Don and {He}, Jiansen and {Hou}, Junfeng and {Hou}, Zhenyong and {Jin}, Chunlan and {Li}, Wenya and {Lin}, Jiaben and {Nandy}, Dibyendu and {Pant}, Vaibhav and {Romoli}, Marco and {Sakao}, Taro and {Krishna Prasad}, Sayamanthula and {Shen}, Fang and {Su}, Yang and {Toriumi}, Shin and {Tripathi}, Durgesh and {Wang}, Linghua and {Wang}, Jingjing and {Xia}, Lidong and {Xiong}, Ming and {Yan}, Yihua and {Yang}, Liping and {Yang}, Shangbin and {Zhang}, Mei and {Zhou}, Guiping and {Zhu}, Xiaoshuai and {Wang}, Jingxiu and {Wang}, Chi},
        title = "{Probing Solar Polar Regions}",
      journal = {Chinese Journal of Space Science},
         year = 2025,
        month = jul,
       volume = {45},
       number = {4},
        pages = {913-942},
          doi = {10.11728/cjss2025.04.2025-0054},
archivePrefix = {arXiv},
       eprint = {2506.20502},
 primaryClass = {astro-ph.SR},
       adsurl = {https://ui.adsabs.harvard.edu/abs/2025ChJSS..45..913D}
}

@ARTICLE{gie23,
       author = {{Gieseler}, Jan and {Dresing}, Nina and {Palmroos}, Christian and {Freiherr von Forstner}, Johan L. and {Price}, Daniel J. and {Vainio}, Rami and {Kouloumvakos}, Athanasios and {Rodr{\'\i}guez-Garc{\'\i}a}, Laura and {Trotta}, Domenico and {G{\'e}not}, Vincent and {Masson}, Arnaud and {Roth}, Markus and {Veronig}, Astrid},
        title = "{Solar-MACH: An open-source tool to analyze solar magnetic connection configurations}",
      journal = {Frontiers in Astronomy and Space Sciences},
         year = 2023,
        month = feb,
       volume = {9},
          eid = {384},
        pages = {384},
          doi = {10.3389/fspas.2022.1058810},
archivePrefix = {arXiv},
       eprint = {2210.00819},
 primaryClass = {astro-ph.SR},
       adsurl = {https://ui.adsabs.harvard.edu/abs/2023FrASS...958810G}
}

@ARTICLE{gan23,
       author = {{Gan}, Weiqun and {Zhu}, Cheng and {Deng}, Yuanyong and {Zhang}, Zhe and {Chen}, Bo and {Huang}, Yu and {Deng}, Lei and {Wu}, Haiyan and {Zhang}, Haiying and {Li}, Hui and {Su}, Yang and {Su}, Jiangtao and {Feng}, Li and {Wu}, Jian and {Cui}, Jijun and {Wang}, Chi and {Chang}, Jin and {Yin}, Zengshan and {Xiong}, Weiming and {Chen}, Bin and {Yang}, Jianfeng and {Li}, Fu and {Lin}, Jiaben and {Hou}, Junfeng and {Bai}, Xianyong and {Chen}, Dengyi and {Zhang}, Yan and {Hu}, Yiming and {Liang}, Yaoming and {Wang}, Jianping and {Song}, Kefei and {Guo}, Quanfeng and {He}, Lingping and {Zhang}, Guang and {Wang}, Peng and {Bao}, Haicao and {Cao}, Caixia and {Bai}, Yanping and {Chen}, Binglong and {He}, Tao and {Li}, Xinyu and {Zhang}, Ye and {Liao}, Xing and {Jiang}, Hu and {Li}, Youping and {Su}, Yingna and {Lei}, Shijun and {Chen}, Wei and {Li}, Ying and {Zhao}, Jie and {Li}, Jingwei and {Ge}, Yunyi and {Zou}, Ziming and {Hu}, Tai and {Su}, Miao and {Ji}, Haidong and {Gu}, Mei and {Zheng}, Yonghuang and {Xu}, Dezhen and {Wang}, Xing},
        title = "{The Advanced Space-Based Solar Observatory (ASO-S)}",
      journal = {\solphys},
         year = 2023,
        month = may,
       volume = {298},
       number = {5},
          eid = {68},
        pages = {68},
          doi = {10.1007/s11207-023-02166-x},
       adsurl = {https://ui.adsabs.harvard.edu/abs/2023SoPh..298...68G}
}

@ARTICLE{go07,
       author = {{Golub}, L. and {DeLuca}, E. and {Austin}, G. and {Bookbinder}, J. and {Caldwell}, D. and {Cheimets}, P. and {Cirtain}, J. and {Cosmo}, M. and {Reid}, P. and {Sette}, A. and {Weber}, M. and {Sakao}, T. and {Kano}, R. and {Shibasaki}, K. and {Hara}, H. and {Tsuneta}, S. and {Kumagai}, K. and {Tamura}, T. and {Shimojo}, M. and {McCracken}, J. and {Carpenter}, J. and {Haight}, H. and {Siler}, R. and {Wright}, E. and {Tucker}, J. and {Rutledge}, H. and {Barbera}, M. and {Peres}, G. and {Varisco}, S.},
        title = "{The X-Ray Telescope (XRT) for the Hinode Mission}",
      journal = {\solphys},
         year = 2007,
        month = jun,
       volume = {243},
       number = {1},
        pages = {63-86},
          doi = {10.1007/s11207-007-0182-1},
       adsurl = {https://ui.adsabs.harvard.edu/abs/2007SoPh..243...63G}
}

@ARTICLE{ko07,
       author = {{Kosugi}, T. and {Matsuzaki}, K. and {Sakao}, T. and {Shimizu}, T. and {Sone}, Y. and {Tachikawa}, S. and {Hashimoto}, T. and {Minesugi}, K. and {Ohnishi}, A. and {Yamada}, T. and {Tsuneta}, S. and {Hara}, H. and {Ichimoto}, K. and {Suematsu}, Y. and {Shimojo}, M. and {Watanabe}, T. and {Shimada}, S. and {Davis}, J.~M. and {Hill}, L.~D. and {Owens}, J.~K. and {Title}, A.~M. and {Culhane}, J.~L. and {Harra}, L.~K. and {Doschek}, G.~A. and {Golub}, L.},
        title = "{The Hinode (Solar-B) Mission: An Overview}",
      journal = {\solphys},
         year = 2007,
        month = jun,
       volume = {243},
       number = {1},
        pages = {3-17},
          doi = {10.1007/s11207-007-9014-6},
       adsurl = {https://ui.adsabs.harvard.edu/abs/2007SoPh..243....3K}
}

@ARTICLE{sav10,
       author = {{Savani}, N.~P. and {Owens}, M.~J. and {Rouillard}, A.~P. and {Forsyth}, R.~J. and {Davies}, J.~A.},
        title = "{Observational Evidence of a Coronal Mass Ejection Distortion Directly Attributable to a Structured Solar Wind}",
      journal = {\apjl},
         year = 2010,
        month = may,
       volume = {714},
       number = {1},
        pages = {L128-L132},
          doi = {10.1088/2041-8205/714/1/L128},
       adsurl = {https://ui.adsabs.harvard.edu/abs/2010ApJ...714L.128S}
}

@ARTICLE{pla16,
       author = {{Plainaki}, Christina and {Lilensten}, Jean and {Radioti}, Aikaterini and {Andriopoulou}, Maria and {Milillo}, Anna and {Nordheim}, Tom A. and {Dandouras}, Iannis and {Coustenis}, Athena and {Grassi}, Davide and {Mangano}, Valeria and {Massetti}, Stefano and {Orsini}, Stefano and {Lucchetti}, Alice},
        title = "{Planetary space weather: scientific aspects and future perspectives}",
      journal = {Journal of Space Weather and Space Climate},
         year = 2016,
        month = aug,
       volume = {6},
          eid = {A31},
        pages = {A31},
          doi = {10.1051/swsc/2016024},
       adsurl = {https://ui.adsabs.harvard.edu/abs/2016JSWSC...6A..31P}
}

@ARTICLE{xia14,
       author = {{Xia}, C. and {Keppens}, R. and {Antolin}, P. and {Porth}, O.},
        title = "{Simulating the in Situ Condensation Process of Solar Prominences}",
      journal = {\apjl},
         year = 2014,
        month = sep,
       volume = {792},
       number = {2},
          eid = {L38},
        pages = {L38},
          doi = {10.1088/2041-8205/792/2/L38},
archivePrefix = {arXiv},
       eprint = {1408.4249},
 primaryClass = {astro-ph.SR},
       adsurl = {https://ui.adsabs.harvard.edu/abs/2014ApJ...792L..38X}
}

@ARTICLE{dai21,
       author = {{Dai}, Jun and {Zhang}, Qingmin and {Zhang}, Yanjie and {Xu}, Zhe and {Su}, Yingna and {Ji}, Haisheng},
        title = "{Oscillations and Mass Draining that Lead to a Sympathetic Eruption of a Quiescent Filament}",
      journal = {\apj},
         year = 2021,
        month = dec,
       volume = {923},
       number = {1},
          eid = {74},
        pages = {74},
          doi = {10.3847/1538-4357/ac2d97},
archivePrefix = {arXiv},
       eprint = {2110.04695},
 primaryClass = {astro-ph.SR},
       adsurl = {https://ui.adsabs.harvard.edu/abs/2021ApJ...923...74D}
}

@ARTICLE{par14,
       author = {{Parenti}, Susanna},
        title = "{Solar Prominences: Observations}",
      journal = {Living Reviews in Solar Physics},
         year = 2014,
        month = dec,
       volume = {11},
       number = {1},
          eid = {1},
        pages = {1},
          doi = {10.12942/lrsp-2014-1},
       adsurl = {https://ui.adsabs.harvard.edu/abs/2014LRSP...11....1P}
}

@ARTICLE{for06,
       author = {{Forbes}, T.~G. and {Linker}, J.~A. and {Chen}, J. and {Cid}, C. and {K{\'o}ta}, J. and {Lee}, M.~A. and {Mann}, G. and {Miki{\'c}}, Z. and {Potgieter}, M.~S. and {Schmidt}, J.~M. and {Siscoe}, G.~L. and {Vainio}, R. and {Antiochos}, S.~K. and {Riley}, P.},
        title = "{CME Theory and Models}",
      journal = {\ssr},
         year = 2006,
        month = mar,
       volume = {123},
       number = {1-3},
        pages = {251-302},
          doi = {10.1007/s11214-006-9019-8},
       adsurl = {https://ui.adsabs.harvard.edu/abs/2006SSRv..123..251F}
}

@ARTICLE{in06,
       author = {{Inhester}, Bernd},
        title = "{Stereoscopy basics for the STEREO mission}",
      journal = {arXiv e-prints},
         year = 2006,
        month = dec,
          eid = {astro-ph/0612649},
        pages = {astro-ph/0612649},
          doi = {10.48550/arXiv.astro-ph/0612649},
archivePrefix = {arXiv},
       eprint = {astro-ph/0612649},
 primaryClass = {astro-ph},
       adsurl = {https://ui.adsabs.harvard.edu/abs/2006astro.ph.12649I}
}

@ARTICLE{bem09,
       author = {{Bemporad}, A.},
        title = "{Stereoscopic Reconstruction from STEREO/EUV Imagers Data of the Three-dimensional Shape and Expansion of an Erupting Prominence}",
      journal = {\apj},
         year = 2009,
        month = aug,
       volume = {701},
       number = {1},
        pages = {298-305},
          doi = {10.1088/0004-637X/701/1/298},
       adsurl = {https://ui.adsabs.harvard.edu/abs/2009ApJ...701..298B}
}

@ARTICLE{lie09,
       author = {{Liewer}, P.~C. and {De Jong}, E.~M. and {Hall}, J.~R. and {Howard}, R.~A. and {Thompson}, W.~T. and {Culhane}, J.~L. and {Bone}, L. and {van Driel-Gesztelyi}, L.},
        title = "{Stereoscopic Analysis of the 19 May 2007 Erupting Filament}",
      journal = {\solphys},
         year = 2009,
        month = may,
       volume = {256},
       number = {1-2},
        pages = {57-72},
          doi = {10.1007/s11207-009-9363-4},
archivePrefix = {arXiv},
       eprint = {0904.1055},
 primaryClass = {astro-ph.SR},
       adsurl = {https://ui.adsabs.harvard.edu/abs/2009SoPh..256...57L}
}

@ARTICLE{lie11,
       author = {{Liewer}, P.~C. and {Hall}, J.~R. and {Howard}, R.~A. and {De Jong}, E.~M. and {Thompson}, W.~T. and {Thernisien}, A.},
        title = "{Stereoscopic analysis of STEREO/SECCHI data for CME trajectory determination}",
      journal = {Journal of Atmospheric and Solar-Terrestrial Physics},
         year = 2011,
        month = jun,
       volume = {73},
       number = {10},
        pages = {1173-1186},
          doi = {10.1016/j.jastp.2010.09.004},
       adsurl = {https://ui.adsabs.harvard.edu/abs/2011JASTP..73.1173L}
}

@ARTICLE{tho06,
       author = {{Thompson}, W.~T.},
        title = "{Coordinate systems for solar image data}",
      journal = {\aap},
         year = 2006,
        month = apr,
       volume = {449},
       number = {2},
        pages = {791-803},
          doi = {10.1051/0004-6361:20054262},
       adsurl = {https://ui.adsabs.harvard.edu/abs/2006A&A...449..791T}
}

@ARTICLE{tho12,
       author = {{Thompson}, W.~T. and {Kliem}, B. and {T{\"o}r{\"o}k}, T.},
        title = "{3D Reconstruction of a Rotating Erupting Prominence}",
      journal = {\solphys},
         year = 2012,
        month = feb,
       volume = {276},
       number = {1-2},
        pages = {241-259},
          doi = {10.1007/s11207-011-9868-5},
archivePrefix = {arXiv},
       eprint = {1112.3388},
 primaryClass = {astro-ph.SR},
       adsurl = {https://ui.adsabs.harvard.edu/abs/2012SoPh..276..241T}
}

@ARTICLE{feng12,
       author = {{Feng}, L. and {Inhester}, B. and {Wei}, Y. and {Gan}, W.~Q. and {Zhang}, T.~L. and {Wang}, M.~Y.},
        title = "{Morphological Evolution of a Three-dimensional Coronal Mass Ejection Cloud Reconstructed from Three Viewpoints}",
      journal = {\apj},
         year = 2012,
        month = may,
       volume = {751},
       number = {1},
          eid = {18},
        pages = {18},
          doi = {10.1088/0004-637X/751/1/18},
archivePrefix = {arXiv},
       eprint = {1203.3261},
 primaryClass = {astro-ph.SR},
       adsurl = {https://ui.adsabs.harvard.edu/abs/2012ApJ...751...18F}
}

@ARTICLE{mich03,
       author = {{Micha{\l}ek}, G. and {Gopalswamy}, N. and {Yashiro}, S.},
        title = "{A New Method for Estimating Widths, Velocities, and Source Location of Halo Coronal Mass Ejections}",
      journal = {\apj},
         year = 2003,
        month = feb,
       volume = {584},
       number = {1},
        pages = {472-478},
          doi = {10.1086/345526},
archivePrefix = {arXiv},
       eprint = {0710.4524},
 primaryClass = {astro-ph},
       adsurl = {https://ui.adsabs.harvard.edu/abs/2003ApJ...584..472M}
}

@ARTICLE{the06,
       author = {{Thernisien}, A.~F.~R. and {Howard}, R.~A. and {Vourlidas}, A.},
        title = "{Modeling of Flux Rope Coronal Mass Ejections}",
      journal = {\apj},
         year = 2006,
        month = nov,
       volume = {652},
       number = {1},
        pages = {763-773},
          doi = {10.1086/508254},
       adsurl = {https://ui.adsabs.harvard.edu/abs/2006ApJ...652..763T}
}
\bibliographystyle{aasjournalv7}

\end{document}